%% file: main.tex
\documentclass[%
 reprint,
 superscriptaddress,
 amsmath,amssymb,
 aps,
 prb,
 floatfix,
]{revtex4-1}

\usepackage{graphicx}
\usepackage{dcolumn}
\usepackage{bm}
\usepackage{overpic}
\usepackage{multirow}
\usepackage{placeins}

\begin{document}

\title{H$_2$-Like Units and Pressure-Tunable High-$T_\mathrm{c}$ Superconductivity in NaH$_6$}

\author{Tianran Chen}\email{tchen34@utk.edu}\affiliation{Department of Physics and Astronomy, University of Tennessee, Knoxville, Tennessee 37996, USA}
\author{Taner Yildirim}\email{taner@nist.gov}\affiliation{NIST Center for Neutron Research, National Institute of Standards and Technology,
 Gaithersburg, MD 20899-8562, USA}

\date{\today}

\begin{abstract}

Hydrogen-rich materials provide a promising route toward high-temperature
conventional superconductivity, although the extreme pressures required for
most predicted and experimentally realized hydride superconductors remain a
major challenge. Here, we present a comprehensive first-principles study of
NaH$_6$, identified in our large-scale search for high-$T_\mathrm{c}$ hydrides and
independently predicted in a subsequent high-throughput study. We find a rich
structural landscape involving several closely competing phases, with
$R\bar{3}m$ and $Pm\bar{3}m$ NaH$_6$ dominating over a broad pressure range.
Remarkably, $R\bar{3}m$ NaH$_6$ remains dynamically stable down to 10~GPa,
where its calculated $T_\mathrm{c}$ is 34~K. Its $T_\mathrm{c}$ increases rapidly with
pressure, reaching 99~K at 50~GPa.
The cubic $Pm\bar{3}m$ phase becomes thermodynamically competitive at higher pressures and exhibits an exceptionally high calculated
$T_\mathrm{c}=230$~K already at 50~GPa, decreasing only weakly upon further
compression. Molecular-dynamics simulations and phonon-renormalization
calculations support structural persistence at finite temperature on the
sampled simulation timescale and indicate that anharmonic effects on their lattice dynamics are
relatively small. Despite their high transition temperatures, both phases
contain elongated H$_2$-like units, conventionally considered unfavorable
for high-$T_\mathrm{c}$ hydride superconductivity. Analysis of the mode-resolved
electron--phonon coupling reveals a strong connection between Raman-active
hydrogen vibrations and the modes responsible for superconductivity. In
$R\bar{3}m$ NaH$_6$, the linewidths of these modes increase with the tilt of
the H$_2$ units relative to the threefold axis, which grows with pressure,
in parallel with the rapid increase of $T_\mathrm{c}$. A minimal
symmetry-based model built on the vibration-induced change of the
second-moment tensor of the hydrogen sublattice shows that the observed trends are consistent with a significant
contribution from hydrogen-sublattice response components perpendicular
to the threefold axis. These results identify NaH$_6$ as an
unusual molecular-like hydride in which structural reorganization of the
hydrogen sublattice appears to provide an effective route to
high-temperature superconductivity at comparatively moderate pressures.

\end{abstract}

\maketitle

\section{\label{sec:introduction}Introduction}

Hydrogen-rich materials provide a promising route toward
high-temperature conventional superconductivity because the light mass
of hydrogen gives rise to high-frequency lattice vibrations, while
strong coupling between hydrogen-derived electronic states and these
vibrations can produce large electron--phonon coupling (EPC).
This concept has received remarkable support from the discovery of
superconductivity in compressed hydrogen-rich materials. In particular,
H$_3$S and LaH$_{10}$ exhibit superconducting transition temperatures
exceeding 200~K under megabar pressures
\cite{duan2014pressure,errea2015high,liu2017potential,somayazulu2019evidence,drozdov2015conventional}.
These discoveries also demonstrated the predictive power of
first-principles calculations: high-pressure crystal-structure searches
combined with electronic, phonon, and EPC calculations can identify
promising superconducting phases and estimate their transition
temperatures before or alongside experimental measurements. A major
remaining challenge, however, is to realize comparably high transition
temperatures at substantially lower pressures.

An important structural characteristic of many high-$T_\mathrm{c}$ hydrides is
an extended hydrogen sublattice in which the hydrogen atoms are no
longer bound into well-defined H$_2$ molecules. In contrast,
hydrogen-rich structures containing H$_2$-like units have often been
regarded as less favorable for high-temperature superconductivity,
because strong H--H bonding tends to place the corresponding bonding
states well below the Fermi level \cite{sun2019route}. Nevertheless,
the distinction between molecular and extended hydrogen networks becomes
less clear under compression, where H--H distances, intermolecular
interactions, charge transfer, and molecular orientations can evolve
strongly with pressure. Understanding whether H$_2$-like units can
participate effectively in electron--phonon coupling, and whether their
structural degrees of freedom can promote rather than suppress
superconductivity, is therefore an important question in the search for
high-$T_\mathrm{c}$ hydrides.

Most theoretical studies of high-$T_\mathrm{c}$ binary hydrides have focused on
compounds of alkaline-earth, rare-earth, and related metals. Examples
include MgH$_6$, CaH$_6$, YH$_6$, YH$_{10}$, and LaH$_{10}$, for which
high transition temperatures are predicted or observed predominantly
at very high pressures
\cite{feng2015compressed,wang2012superconductive,heil2019superconductivity,
liu2017potential}.
Alkali-metal hydrides have received comparatively less attention as
potential high-$T_\mathrm{c}$ superconductors. Earlier structure searches
established a rich high-pressure sodium--hydrogen phase diagram and
predicted the stability of several sodium polyhydrides, including
NaH$_6$ \cite{baettig2011pressure}. More recently, NaH$_6$ was
independently identified as a promising superconducting hydride in
large-scale computational searches. In our systematic search of more
than 100,000 binary hydride structures, NaH$_6$ emerged as one of the
most promising candidates \cite{ChenYildirimSearch}, while an
independent high-throughput study also predicted a very high
superconducting transition temperature for NaH$_6$
\cite{Shipley2021}. These screening studies, however, leave open the
underlying structural and microscopic physics of superconductivity in
this unusual hydrogen-rich compound.

Here, we present a comprehensive first-principles investigation of the
structural, vibrational, and superconducting properties of NaH$_6$ over
a broad pressure range. We find a remarkably rich energy landscape
containing several closely competing structures, dominated by the
rhombohedral $R\bar{3}m$ and cubic $Pm\bar{3}m$ phases.
$R\bar{3}m$ NaH$_6$ remains dynamically stable down to 10~GPa and its
calculated $T_\mathrm{c}$ is 34~K; remarkably, its transition temperature
increases rapidly with compression, reaching 99~K at 50~GPa. The
$Pm\bar{3}m$ phase is dynamically stable down to 50~GPa and has a
calculated $T_\mathrm{c}$ as high as 230~K at this pressure.
Molecular-dynamics and phonon-renormalization calculations further support
structural persistence at finite temperature on the sampled timescale and
indicate relatively weak anharmonic effects.
Both phases contain distinct H$_2$-like units with elongated H--H
bonds, providing an unusual example in which molecular-like hydrogen
motifs coexist with strong electron--phonon coupling and very high
$T_\mathrm{c}$. By analyzing the mode-resolved EPC, we uncover a strong
correlation between Raman-active hydrogen vibrations and EPC-active
modes. In $R\bar{3}m$ NaH$_6$, the linewidths of the Raman-active modes grow
with the tilt of the H$_2$-like units, which increases with pressure, and a
minimal symmetry-based model of the hydrogen sublattice indicates that this
growth is consistent with a significant contribution from the tensor
components perpendicular to the threefold axis. This provides insight into
how molecular-like hydrogen substructures can participate in
high-temperature superconductivity.

\section{Results and Discussion}

\subsection{Crystal Structures and Stability of NaH$_6$}

Using the USPEX \cite{glass2006uspex} evolutionary algorithm, we searched several thousand
binary sodium hydride Na$_m$H$_n$ structures at 200~GPa. Almost all
low-enthalpy hydrogen-rich ($n>m$) compounds contain H$_2$-like units.
For comparison, we performed similar searches for Mg$_m$H$_n$ and
Al$_m$H$_n$ hydrides and found many low-enthalpy structures that do
not contain H$_2$-like units (see Supplementary Material for details).
This comparison suggests that the formation of H$_2$-like units may
play an important role in stabilizing hydrogen-rich Na$_m$H$_n$
compounds. Structures containing H$_2$-like units have generally been
considered less favorable for high-$T_\mathrm{c}$ superconductivity, which may
partly explain why sodium hydrides have received relatively little
attention as potential high-$T_\mathrm{c}$ superconductors.

During preliminary EPC calculations using coarse k- and q-point grids
for the low-enthalpy metallic sodium hydrides, we found that
$P\bar{1}$ NaH$_6$ and $P2/m$ NaH$_6$ exhibit substantially larger
EPC constants and superconducting transition temperatures than the
other structures examined. Upon further structural optimization, both
structures converge to the higher-symmetry $Pm\bar{3}m$ phase.
The $Pm\bar{3}m$ structure of NaH$_6$ was previously reported by
Baettig and Zurek \cite{baettig2011pressure} as a stable phase at and
above 150~GPa. It was subsequently identified as a promising
high-$T_\mathrm{c}$ superconductor in large-scale computational searches
\cite{ChenYildirimSearch,Shipley2021}.

\begin{figure}[htp]
\centering
\begin{overpic}[width=0.48\textwidth]{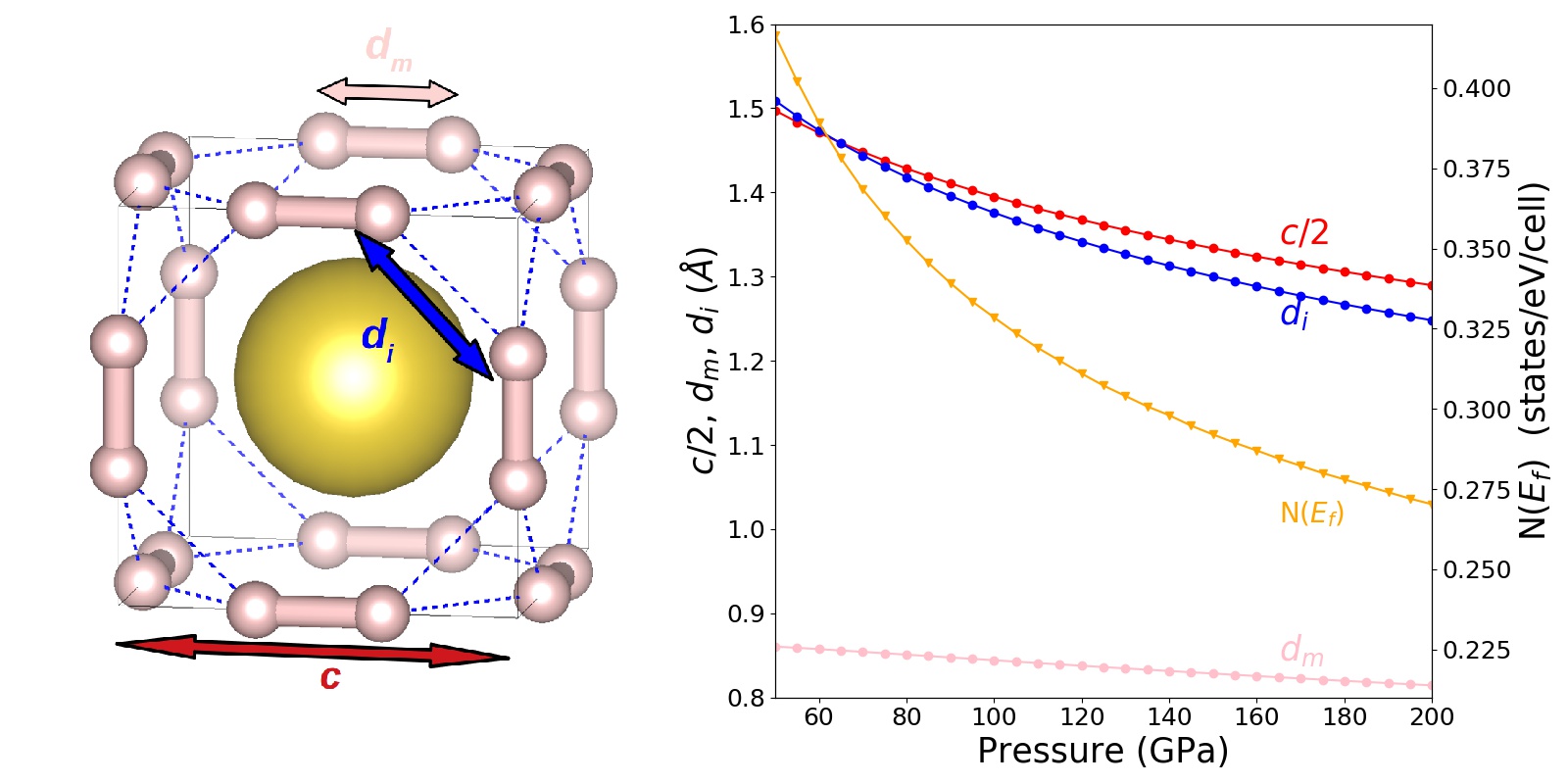}
\put (2,44) {\normalsize\textbf a}
\put (41,44) {\normalsize\textbf b}
\end{overpic}
 \caption{
(a) Crystal structure of $Pm\bar{3}m$ NaH$_6$ at 200~GPa. The yellow
and pink spheres represent Na and H atoms, respectively.
(b) Lattice constant $c/2$, electronic density of states at the Fermi
level $N(E_\mathrm{F})$, intramolecular H--H distance $d_\mathrm{m}$, and intermolecular
H--H distance $d_\mathrm{i}$ as functions of pressure.
}
\label{fig-pressure}
\end{figure}

Structural optimizations and phonon calculations were performed over a
range of pressures. Remarkably, $Pm\bar{3}m$ NaH$_6$ remains
dynamically stable down to 50~GPa. Additional USPEX calculations for
Na$_m$H$_n$ at several pressures show that $Pm\bar{3}m$ NaH$_6$ is
energetically stable against decomposition into NaH and H$_2$ above
60~GPa and becomes the lowest-enthalpy NaH$_6$ phase at 200~GPa
(see Supplementary Material for details).

The crystal structure of $Pm\bar{3}m$ NaH$_6$ is shown in
Fig.~\ref{fig-pressure}(a). The hydrogen atoms form distinct
H$_2$-like units, shown as pink sticks, with an H--H distance
$d_\mathrm{m}=0.861$~\AA\ at 50~GPa. This distance decreases by only 5.4\%
upon compression to 200~GPa. In contrast, the intermolecular H--H
distance $d_\mathrm{i}$, shown as blue dashed lines, is 1.509~\AA\ at 50~GPa
and decreases by 17.3\% upon compression to 200~GPa. The substantially
stronger pressure dependence of $d_\mathrm{i}$ indicates that compression
primarily reduces the separation between neighboring H$_2$-like units
rather than the H--H distance within each unit.

\subsection{Superconductivity in $Pm\bar{3}m$ NaH$_6$}

\begin{figure}[htp]
\centering
\begin{overpic}[width=0.48\textwidth]{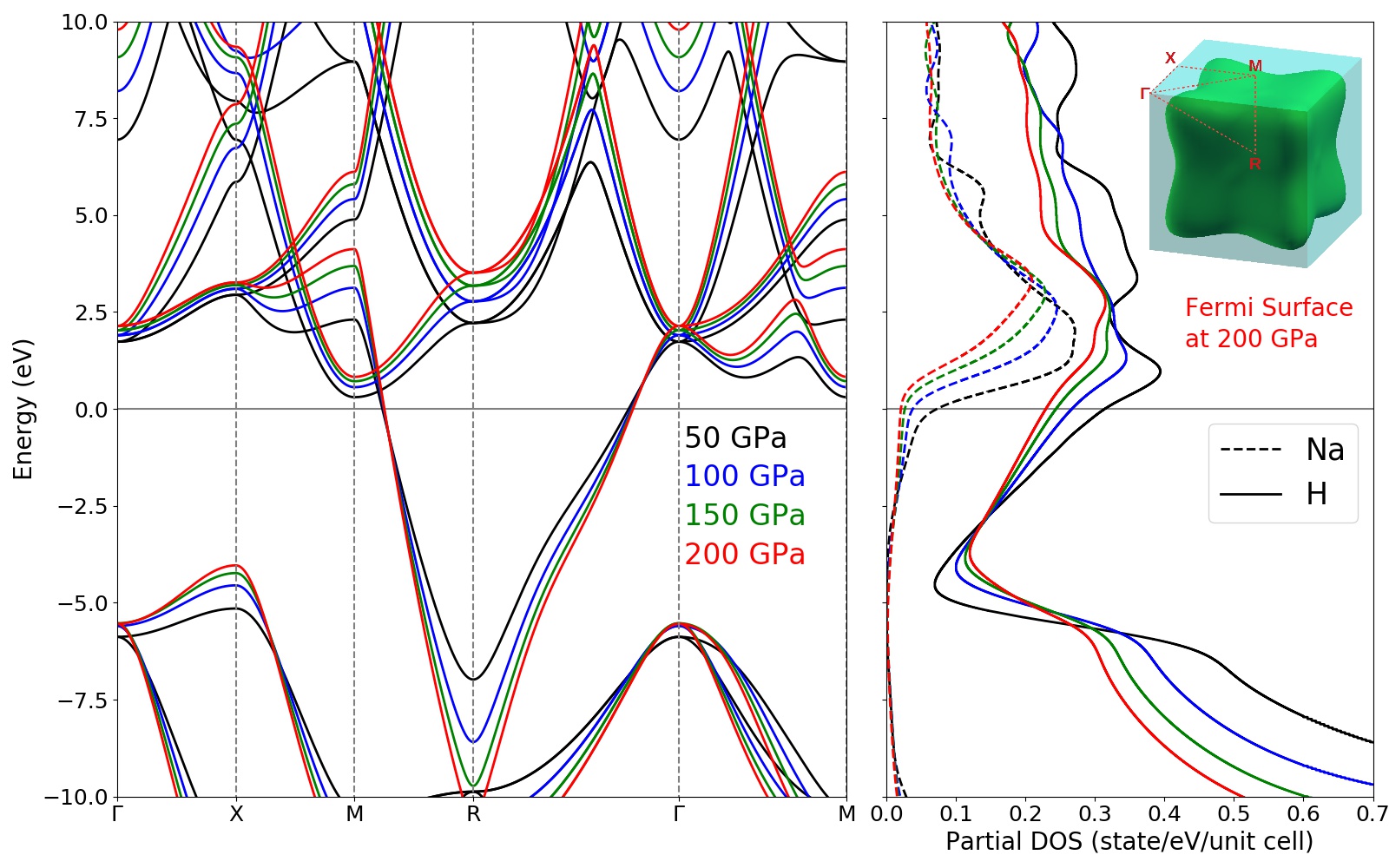}
\put (1,57) {\normalsize\textbf a}
\put (79,57) {\normalsize\textbf b}
\end{overpic}
\caption{
(a) Electronic band structures of $Pm\bar{3}m$ NaH$_6$ at 50, 100,
150, and 200~GPa, shown in black, blue, green, and red, respectively.
The Fermi level is set to zero. (b) Fermi surface of $Pm\bar{3}m$ NaH$_6$ at 200~GPa.
}
\label{fig-eband}
\end{figure}

The electronic band structures and density of states of
$Pm\bar{3}m$ NaH$_6$ at 50, 100, 150, and 200~GPa are shown in
Fig.~\ref{fig-eband}. The electronic structure near the Fermi level
is relatively simple and changes little with pressure, in contrast to
many other hydride superconductors, including H$_3$S
\cite{bianconi2015superconductivity}, YH$_6$ and YH$_{10}$
\cite{heil2019superconductivity}, and CaH$_6$
\cite{wang2012superconductive}. The Fermi surface consists of a single
isolated rounded cube centered at the R point, as shown in
Fig.~\ref{fig-eband}(b). Interestingly, the Fermi level $E_\mathrm{F}$ lies
close to an optimum position for high-$T_\mathrm{c}$ superconductivity. Electron
doping, corresponding to an increase in $E_\mathrm{F}$, would significantly
increase $N_\mathrm{Na}(E_\mathrm{F})$, implying greater charge transfer from H
to Na and, consequently, a reduction in $T_\mathrm{c}$. Conversely, hole
doping would decrease $N_\mathrm{H}(E_\mathrm{F})$ significantly, which is also
expected to reduce $T_\mathrm{c}$.

\begin{figure}[htp]
\centering
\begin{overpic}[width=0.48\textwidth]{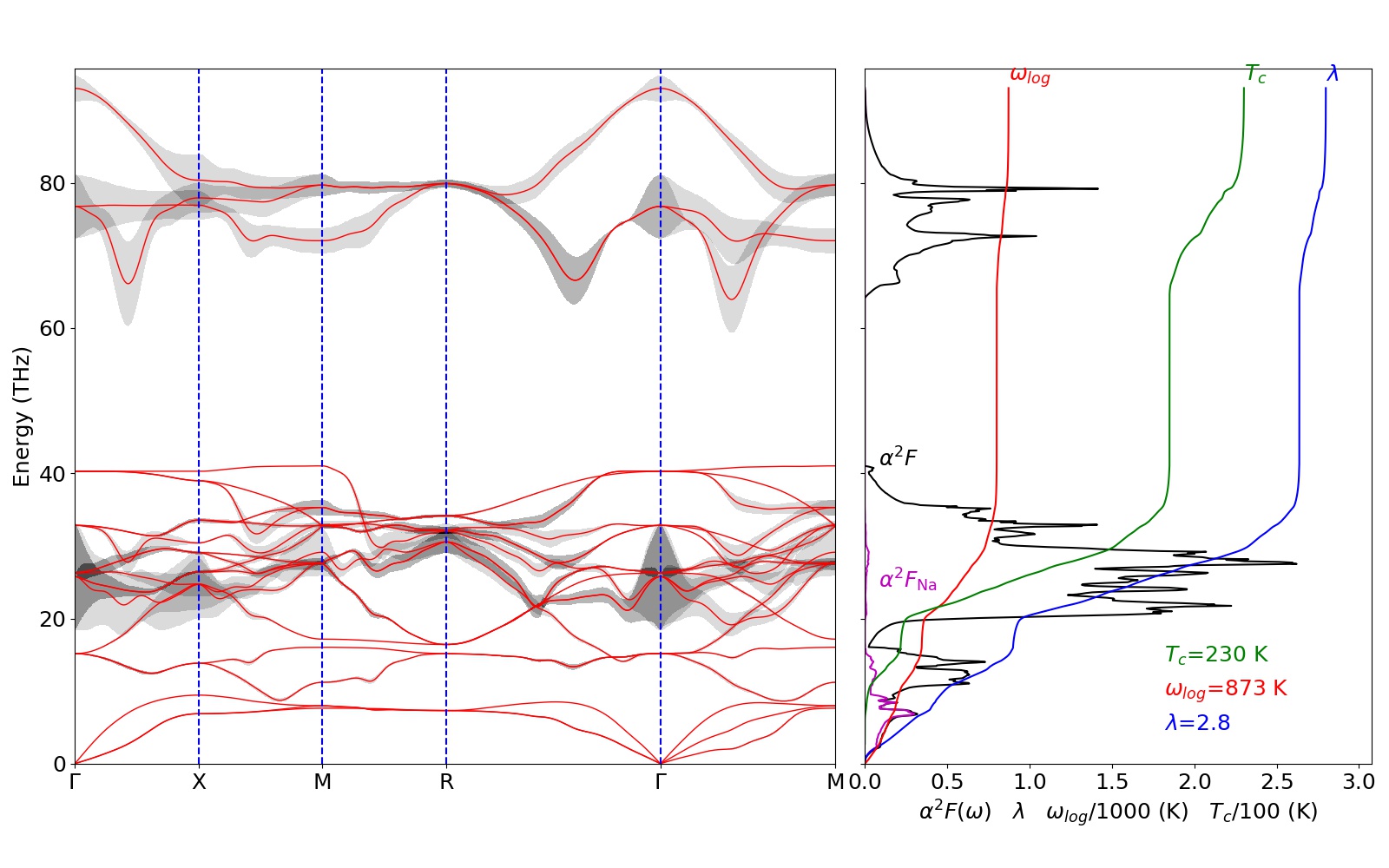}
\put (1,54) {\normalsize\textbf a}
\put (65,54) {\normalsize\textbf b}
\end{overpic}
\caption{
(a) Phonon dispersion of $Pm\bar{3}m$ NaH$_6$ at 50~GPa. The gray
shaded areas represent the phonon linewidths.
(b) Eliashberg spectral function $\alpha^2F$. The black and purple
curves show the total $\alpha^2F$ and the Na-projected contribution,
respectively. The green, red, and blue curves show the integrated
$T_\mathrm{c}$, $\omega_{\mathrm{log}}$, and $\lambda$ as functions of phonon frequency
$\omega$, respectively. Corresponding results at 100, 150, and
200~GPa are provided in the Supplementary Material.
}
\label{fig-a2F}
\end{figure}

High-accuracy phonon and EPC calculations were performed using
density-functional perturbation theory (DFPT)
\cite{baroni2001phonons}, as implemented in QUANTUM ESPRESSO
\cite{giannozzi2009quantum}, at several pressures. The superconducting
transition temperature $T_\mathrm{c}$ was calculated using the Allen--Dynes
modified McMillan formula
\cite{allen1975transition,allen1980dynamical}, with a typical Coulomb
pseudopotential parameter $\mu^*=0.10$.
\begin{equation}
T_\mathrm{c}=\frac{f_1f_2\omega_{\mathrm{log}}}{1.2}
\exp\left[
-\frac{1.04(1+\lambda)}
{\lambda-\mu^*(1+0.62\lambda)}
\right],
\end{equation}
where
\begin{equation}
\omega_{\mathrm{log}}=\exp\left[
\frac{2}{\lambda}
\int_0^{\omega_{\max}}
\frac{\alpha^2F(\omega)}{\omega}\ln(\omega)\,d\omega
\right],
\end{equation}
\begin{equation}
\omega_2=
\left[
\frac{2}{\lambda}
\int_0^{\omega_{\max}}
\alpha^2F(\omega)\omega\,d\omega
\right]^{1/2},
\end{equation}
\begin{equation}
\lambda=
2\int_0^{\omega_{\max}}
\frac{\alpha^2F(\omega)}{\omega}\,d\omega,
\end{equation}
\begin{equation}
f_1=
\left[
1+\left(\frac{\lambda}{2.46(1+3.8\mu^*)}\right)^{3/2}
\right]^{1/3},
\end{equation}
\begin{equation}
f_2=
1+\frac{\lambda^2(\omega_2/\omega_{\mathrm{log}}-1)}
{\lambda^2+
\left[
1.82(1+6.3\mu^*)\omega_2/\omega_{\mathrm{log}}
\right]^2}.
\end{equation}

At 50~GPa, the calculated critical temperature $T_\mathrm{c}$, characteristic
phonon frequency $\omega_{\mathrm{log}}$, and EPC constant $\lambda$ are
230~K, 873~K, and 2.80, respectively. Upon compression to 200~GPa,
$T_\mathrm{c}$ gradually decreases to 215~K, corresponding to a pressure
coefficient $dT_\mathrm{c}/dP=-0.1$~K\,GPa$^{-1}$. The superconducting parameters at
different pressures are listed in the Supplementary Material.
Figure~\ref{fig-a2F} shows the phonon dispersion, phonon linewidths,
and Eliashberg spectral function $\alpha^2F$
\cite{eliashberg1960interactions,eliashberg1961temperature} at
50~GPa; the corresponding results at other pressures are provided in
the Supplementary Material.

At 200~GPa, the acoustic and optical phonon branches are separated by
a gap of approximately 4~THz. This gap decreases with decreasing
pressure and disappears at 50~GPa. The Na-projected contribution to
$\alpha^2F$ is concentrated in the low-frequency acoustic modes and
accounts for less than 5\% of the total $\alpha^2F$. The presence of
H$_2$-like units gives rise to a phonon spectrum that differs
significantly from those of many other high-$T_\mathrm{c}$ hydride
superconductors. At 200~GPa, $\omega_{\mathrm{max}}$ reaches 112~THz, nearly
twice the maximum phonon frequencies of LaH$_{10}$ and H$_3$S at the
same pressure. Over the pressure range 50--200~GPa, the high-frequency
modes associated with stretching of the H$_2$-like units are separated
from the remaining phonon modes by 20.5--22.9~THz and contribute
18.4--19.5\% of the total $\alpha^2F$.

DFPT phonon calculations indicate that $Pm\bar{3}m$ NaH$_6$ is
dynamically stable at and above 50~GPa at zero temperature. However,
dynamical stability at zero temperature does not by itself establish
stability at finite temperature. At 50~GPa, $Pm\bar{3}m$ NaH$_6$ lies
0.08~eV/atom above the convex hull and 0.02~eV/atom above the
NaH + 2.5H$_2$ decomposition line. We therefore performed molecular
dynamics (MD) simulations using VASP \cite{kresse1996software} at
$T=300$~K and pressures of 50 and 200~GPa using a
$5\times5\times5$ supercell. Finite-displacement (FD) phonon
calculations were also performed using VASP and Phonopy
\cite{togo2015first}, with the same supercell and k-point meshes, to
assess possible anharmonic effects, which are known to be significant
in several other hydride superconductors
\cite{errea2015high,errea2016quantum,rousseau2011anharmonicity}.
The lattice parameters were held fixed during the MD simulations.

\begin{figure}[htp]
\centering
\begin{overpic}[width=0.48\textwidth]{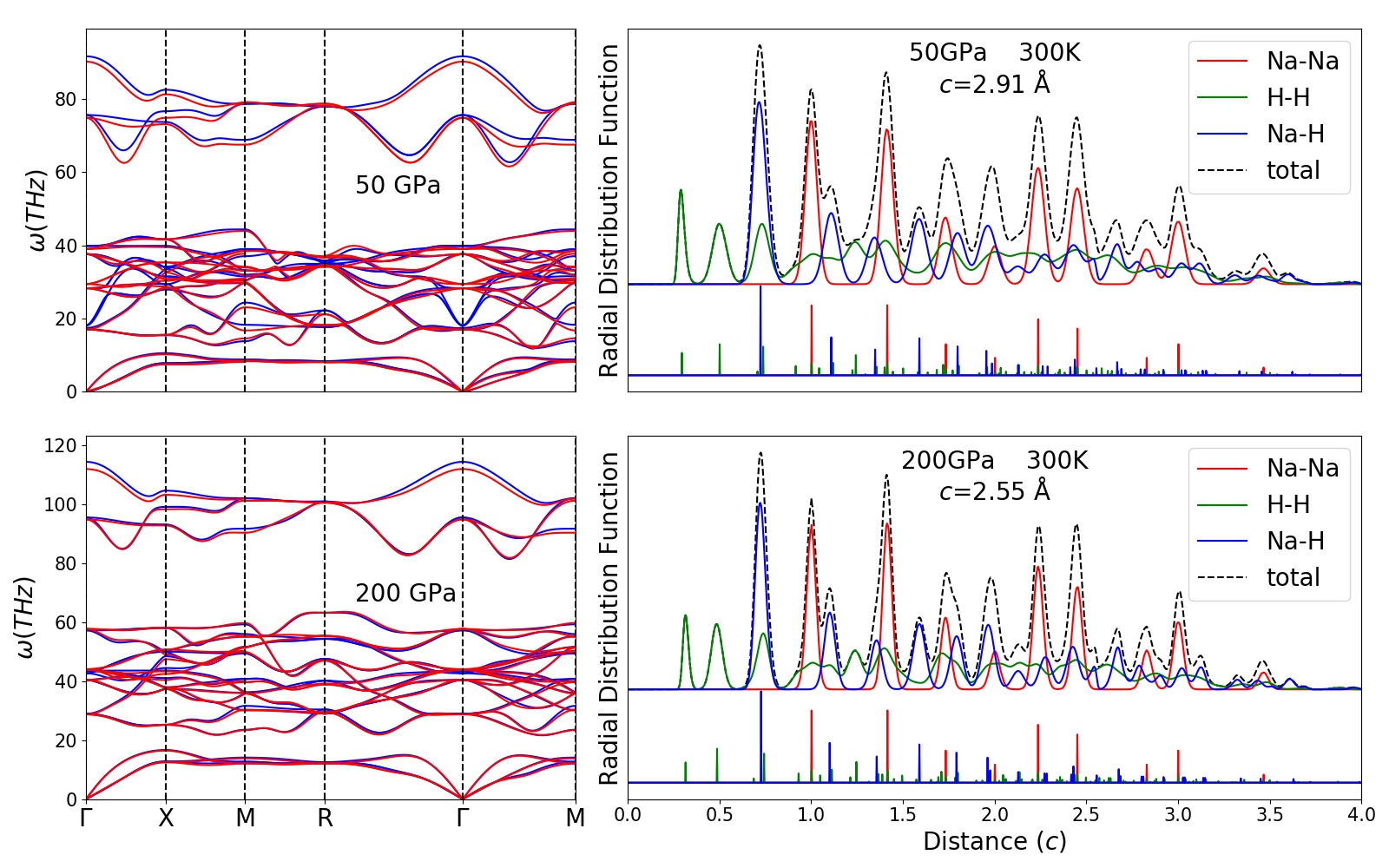}
\put (9,58) {\normalsize\textbf a}
\put (9,28) {\normalsize\textbf b}
\put (46,58) {\normalsize\textbf c}
\put (46,28) {\normalsize\textbf d}
\end{overpic}
 \caption{
Renormalized phonon dispersions (red lines) of $Pm\bar{3}m$ NaH$_6$
obtained from MD simulations using a $5\times5\times5$ supercell at
300~K and (a) 50~GPa and (b) 200~GPa. Harmonic phonon dispersions
calculated using the finite-displacement (FD) method with the same
supercell are shown in blue for comparison. Radial distribution
functions (RDFs) at 300~K and (c) 50~GPa and (d) 200~GPa are shown
as black dashed lines. The red, green, and blue curves represent the
Na--Na, H--H, and Na--H partial RDFs, respectively. The corresponding
RDFs of the static structures are shown below the curves for comparison.
}
\label{fig-renormPh}
\end{figure}

The $Pm\bar{3}m$ structure remains intact in the MD simulations at
both pressures, with no spontaneous decomposition observed over the
several-picosecond simulation time. Phonon quasiparticles were
extracted from the MD trajectories, and the renormalized phonon
frequencies were obtained using DynaPhoPy
\cite{carreras2017dynaphopy}. Figures~\ref{fig-renormPh}(a) and (b)
compare the harmonic and renormalized phonon dispersions at $T=300$~K
and pressures of 50 and 200~GPa. The phonon-frequency shifts are small
and show no clear systematic pressure dependence.
Figures~\ref{fig-renormPh}(c) and (d) show the corresponding radial
distribution functions (RDFs), with radial distances normalized by the
lattice constant $c$. No additional peaks appear relative to the
static zero-temperature RDFs. Together, these results indicate that
$Pm\bar{3}m$ NaH$_6$ remains structurally robust at 300~K on the
timescale of the MD simulations and that anharmonic effects on its
phonon spectrum are relatively weak.

\subsection{Competing Phases and Pressure Evolution of NaH$_6$}

$Pm\bar{3}m$ NaH$_6$ is the lowest-enthalpy phase at 200~GPa.
At 50~GPa, however, we identified 14 additional metallic and
dynamically stable phases of NaH$_6$, many of which have lower
enthalpies than the $Pm\bar{3}m$ phase. Detailed properties of these
phases are provided in the Supplementary Material. Most of the
low-symmetry phases are structurally very similar to either
$R\bar{3}m$ or $Pm\bar{3}m$ NaH$_6$, with structural matching
tolerances of 0.0002--0.015~\AA. The presence of numerous closely
related metastable structures suggests a relatively complex and
shallow structural energy landscape around these two phases.
Table~\ref{table-3phase} compares the physical properties of
$Pm\bar{3}m$, $R\bar{3}m$, and $C2/m$ NaH$_6$ at 50~GPa.
The calculated $T_\mathrm{c}$ values of the $R\bar{3}m$ and $C2/m$ phases
are substantially lower than that of $Pm\bar{3}m$ NaH$_6$.

\begin{table}[htbp]
 \caption{
Physical properties of $Pm\bar{3}m$, $R\bar{3}m$, and $C2/m$
NaH$_6$ at 50~GPa.
}
\label{table-3phase}
\centering
\begin{tabular}{|c|c|c|c|}
\hline
 & $Pm\bar{3}m$ & $R\bar{3}m$ & $C2/m$ \\
\hline
enthalpy (meV) & 0 & $-167.7$ & $-130.2$ \\
volume (\AA$^3$) & 26.79 & 27.49 & 27.47 \\
lattice const (\AA) & 2.992 & 3.072 & 3.027/3.378 \\
lattice angle ($^{\circ}$) & 90 & 78.67 & 109.15/79.48 \\
$T_\mathrm{c}$ (K) & 230 & 99 & 76 \\
\hline
\end{tabular}
\end{table}

\begin{figure}[htp]
\centering
\begin{overpic}[width=0.48\textwidth]{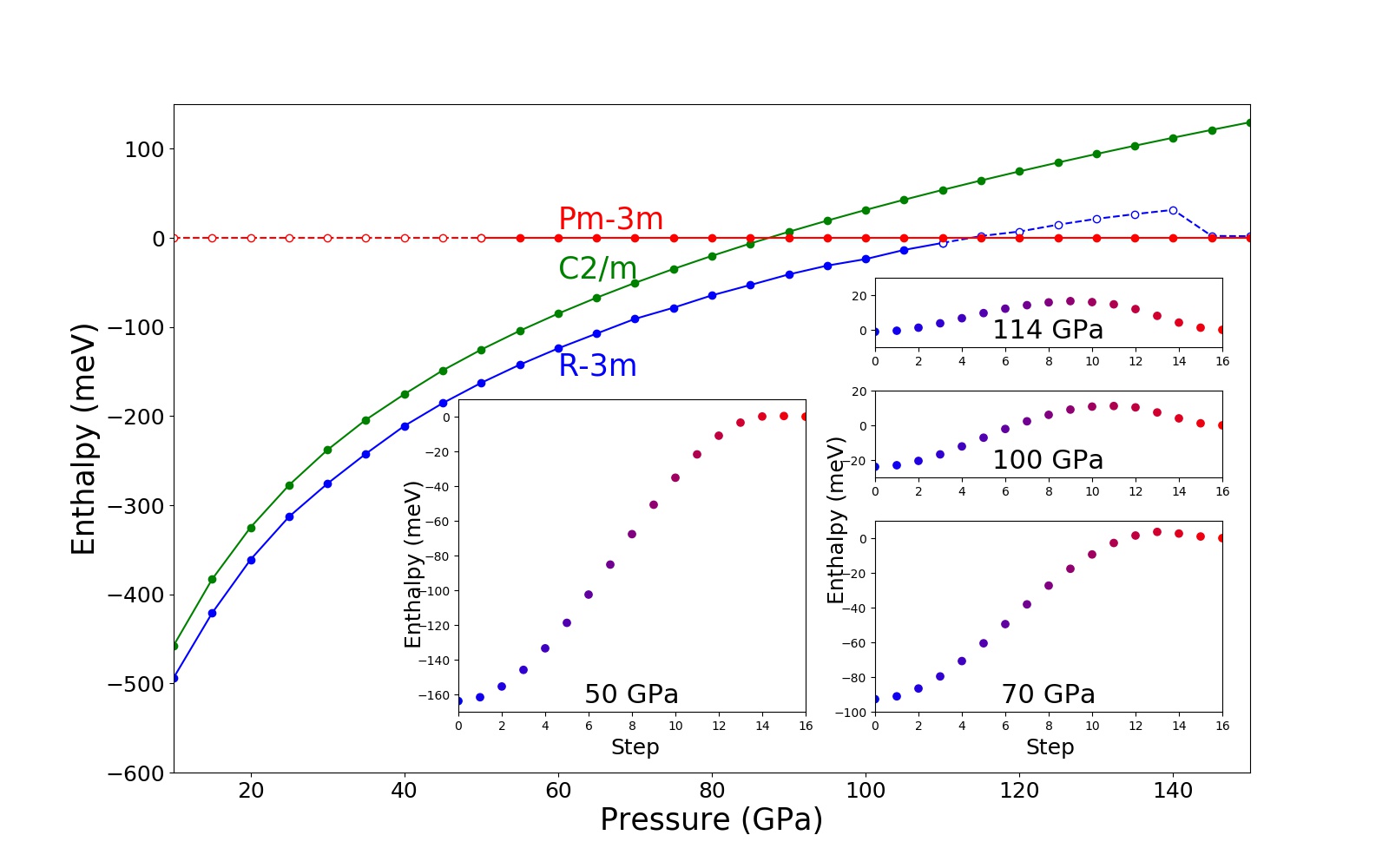}
\put (13,52) {\normalsize\textbf a}
\put (34,30) {\normalsize\textbf b}
\put (64,40) {\normalsize\textbf c}
\put (64,31) {\normalsize\textbf d}
\put (64,21) {\normalsize\textbf e}
\end{overpic}
 \caption{
(a) Enthalpies of $Pm\bar{3}m$, $R\bar{3}m$, and $C2/m$ NaH$_6$
as functions of pressure, shown in red, blue, and green, respectively.
Open circles denote dynamically unstable structures.
(b)--(e) Enthalpy landscapes along the structural transition pathway
between $R\bar{3}m$ and $Pm\bar{3}m$ NaH$_6$ at (b) 50,
(c) 70, (d) 100, and (e) 114~GPa.
}
\label{fig-phaselandscape}
\end{figure}

Further phonon calculations show that both $R\bar{3}m$ and $C2/m$
NaH$_6$ remain dynamically stable down to 10~GPa.
Figure~\ref{fig-phaselandscape} shows the enthalpies of
$Pm\bar{3}m$, $R\bar{3}m$, and $C2/m$ NaH$_6$ as functions of
pressure, calculated using QUANTUM ESPRESSO with
$32\times32\times32$ k-point meshes. At 50~GPa, $Pm\bar{3}m$
NaH$_6$ has the highest enthalpy and the smallest volume per formula
unit among the three phases. The $Pm\bar{3}m$ and $C2/m$ phases
become equal in enthalpy at 88~GPa, while the $Pm\bar{3}m$ and
$R\bar{3}m$ phases become equal in enthalpy at 114~GPa. The
enthalpy of $R\bar{3}m$ NaH$_6$ remains lower than that of the
$C2/m$ phase over the entire pressure range considered.

We performed phonon and EPC calculations for $R\bar{3}m$ NaH$_6$
at 10, 30, 50, 70, 90, and 110~GPa using DFPT with
$16\times16\times16$ k-point and $8\times8\times8$ q-point
meshes. The $R\bar{3}m$ phase remains dynamically stable up to
90~GPa. At 110~GPa, however, small imaginary phonon frequencies
extending to approximately $-2$~THz appear near the $\Gamma$ point.
Finite-displacement phonon calculations using a
$4\times4\times4$ supercell also show these imaginary modes.
We therefore performed MD simulations of $R\bar{3}m$ NaH$_6$ at
110~GPa and 150~K. The structure remains intact during the
simulations, as shown in the Supplementary Material, suggesting
finite-temperature stabilization on the sampled MD timescale.
At and above 140~GPa, the imaginary phonon modes become more
pronounced, and structural relaxation drives $R\bar{3}m$ NaH$_6$
to the $Pm\bar{3}m$ phase. The open circles connected by dashed
lines in Fig.~\ref{fig-phaselandscape} denote dynamically unstable
structures.

\begin{figure}[htp]
\centering
\begin{overpic}[width=0.48\textwidth]{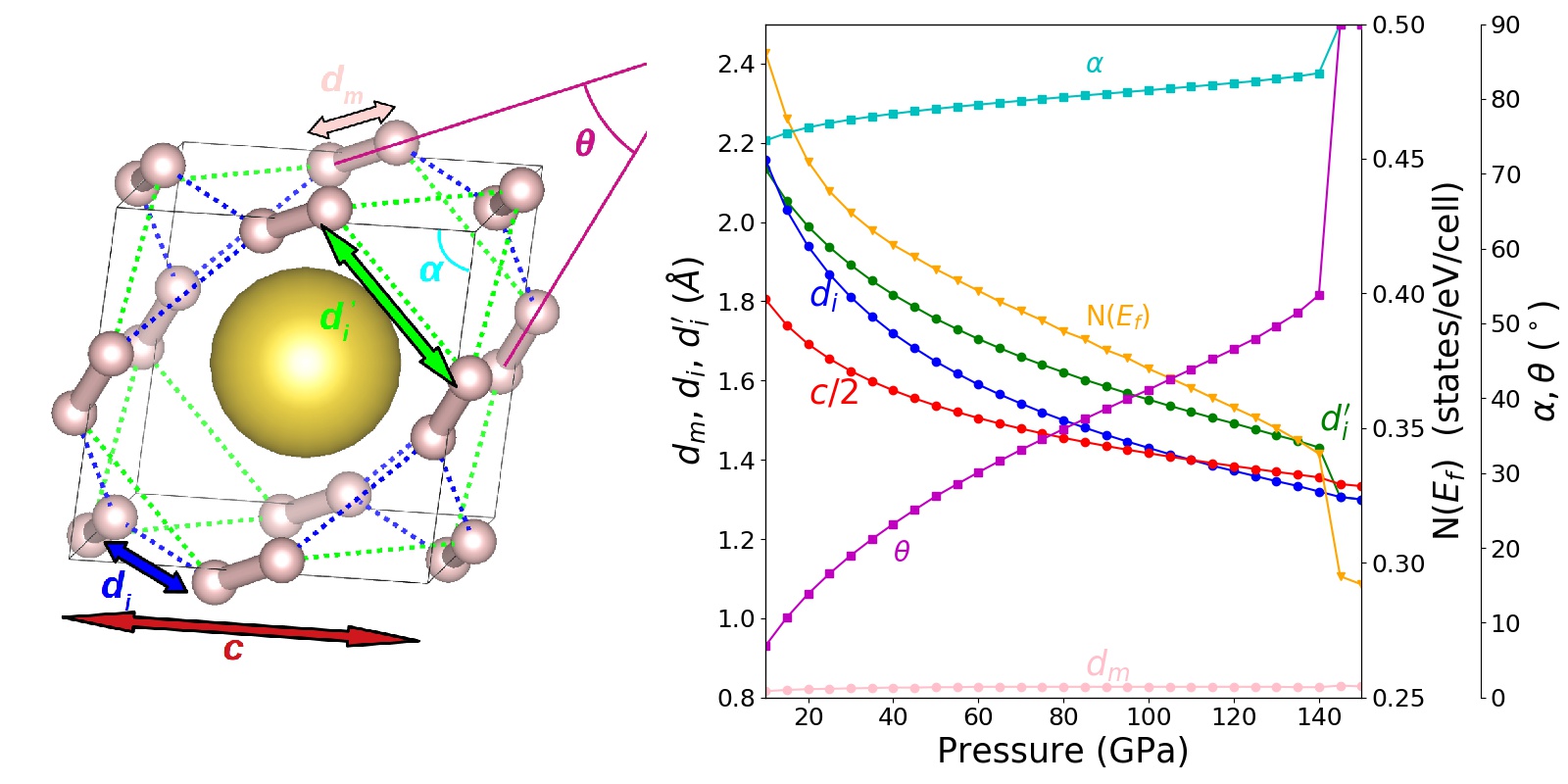}
\put (2,44) {\normalsize\textbf a}
\put (51,44) {\normalsize\textbf b}
\end{overpic}
\caption{
(a) Crystal structure of $R\bar{3}m$ NaH$_6$. The yellow and pink
spheres represent Na and H atoms, respectively. The angles $\alpha$
and $\theta$ denote the angle between the lattice vectors and the
angle between neighboring H$_2$-like units, respectively.
(b) Intramolecular H--H distance $d_\mathrm{m}$, intermolecular H--H distances
$d_\mathrm{i}$ and $d_\mathrm{i}'$, electronic density of states at the Fermi level
$N(E_\mathrm{F})$, lattice angle $\alpha$, and H$_2$-orientation angle
$\theta$ of $R\bar{3}m$ NaH$_6$ over the pressure range
10--150~GPa. Above 140~GPa, structural relaxation transforms the
$R\bar{3}m$ phase into $Pm\bar{3}m$.
}
\label{fig-pressureR-3m}
\end{figure}

To investigate the transition pathway and energy barrier between
$R\bar{3}m$ and $Pm\bar{3}m$ NaH$_6$, we constructed a series of
intermediate structures by assuming that the volume and
$\cos(\alpha)$ vary linearly along the transition pathway, where
$\alpha$ is the angle between the lattice axes of $R\bar{3}m$
NaH$_6$ [see Fig.~\ref{fig-pressureR-3m}(a)]. Lattice parameters
were generated for 15 intermediate structures, and the atomic
positions were optimized at fixed cell parameters using QUANTUM
ESPRESSO with $32\times32\times32$ k-point meshes. As shown in
Figs.~\ref{fig-phaselandscape}(b)--(e), the calculated energy barrier
between the $R\bar{3}m$ and $Pm\bar{3}m$ phases is less than
1~meV at 50~GPa and increases with pressure, reaching 16.5~meV at
114~GPa, where the two phases have equal enthalpies. The relatively
small calculated barrier suggests that the structural transition may
exhibit only modest hysteresis at finite temperature.

\begin{figure}[htp]
\centering
\begin{overpic}[width=0.48\textwidth]{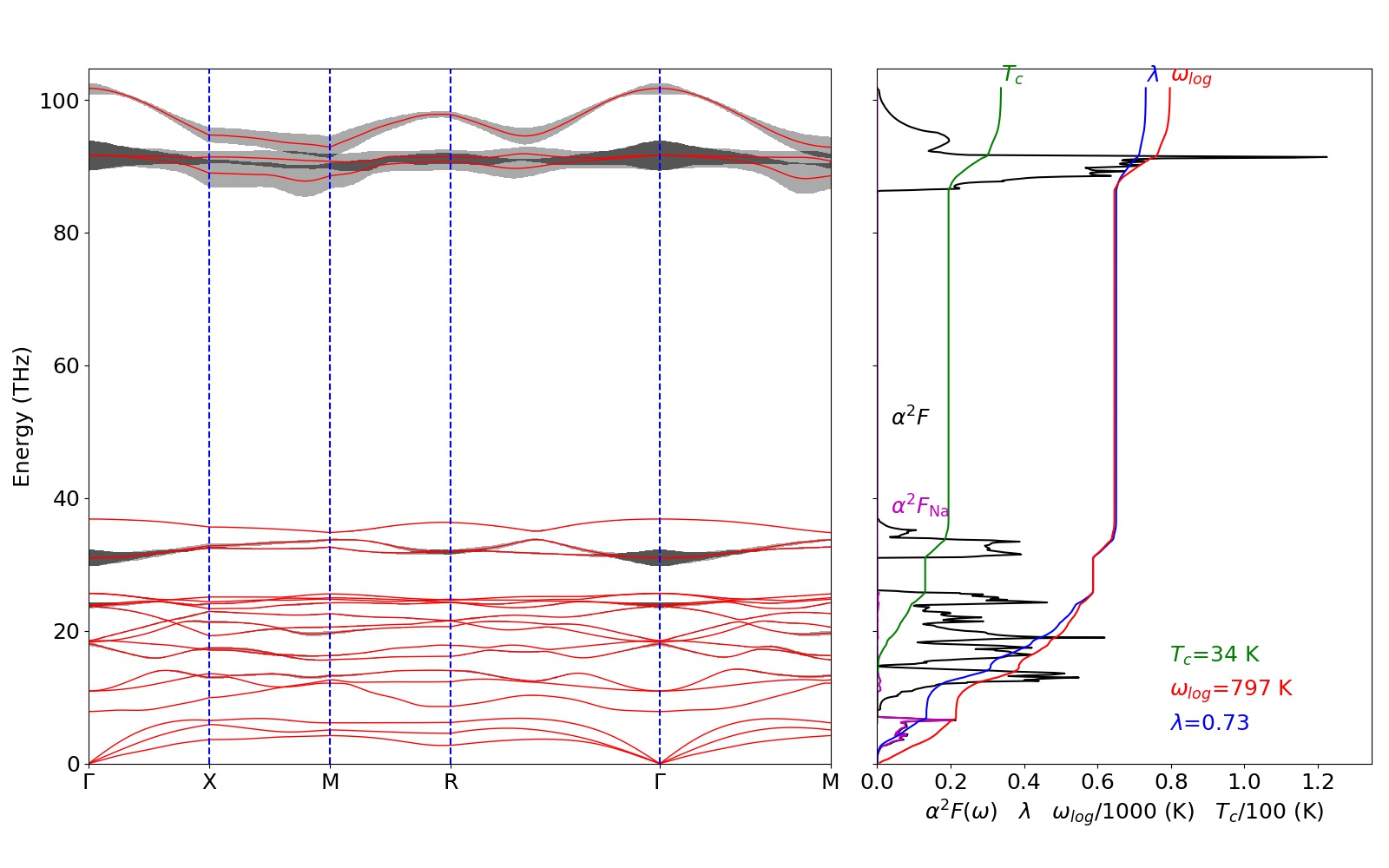}
\put (0,54) {\normalsize\textbf a}
\put (67,54) {\normalsize\textbf b}
\end{overpic}
\caption{
(a) Phonon dispersion of $R\bar{3}m$ NaH$_6$ at 10~GPa. The gray
shaded areas represent the phonon linewidths.
(b) $\alpha^2F$, $\lambda$, $\omega_{\mathrm{log}}$, and $T_\mathrm{c}$ of
$R\bar{3}m$ NaH$_6$ at 10~GPa. The black and purple curves show
the total $\alpha^2F$ and the Na-projected contribution, respectively.
The green, red, and blue curves show the integrated $T_\mathrm{c}$,
$\omega_{\mathrm{log}}$, and $\lambda$ as functions of phonon frequency,
respectively. Corresponding results at 30, 50, 70, and 90~GPa are
provided in the Supplementary Material.
}
\label{fig-a2FR-3m}
\end{figure}

Figure~\ref{fig-a2FR-3m} shows the phonon dispersion, phonon
linewidths, $\alpha^2F$, $\lambda$, $\omega_{\mathrm{log}}$, and $T_\mathrm{c}$ of
$R\bar{3}m$ NaH$_6$ at 10~GPa. Corresponding results at other
pressures are provided in the Supplementary Material. At low
pressures, $R\bar{3}m$ NaH$_6$ exhibits relatively weak
$\alpha^2F$ in the 20--40~THz frequency range, which increases
rapidly upon compression. The resulting pressure coefficient is
$dT_\mathrm{c}/dP=1.6$~K\,GPa$^{-1}$. This large positive pressure dependence is particularly striking because the
electronic band structure near the Fermi level changes only slightly with
compression. A symmetry-based interpretation of this behavior is given in
the following subsection.

\subsection{Symmetry Analysis of the Electron--Phonon Coupling}

\begin{figure}[htp]
\centering
\begin{overpic}[width=0.48\textwidth]{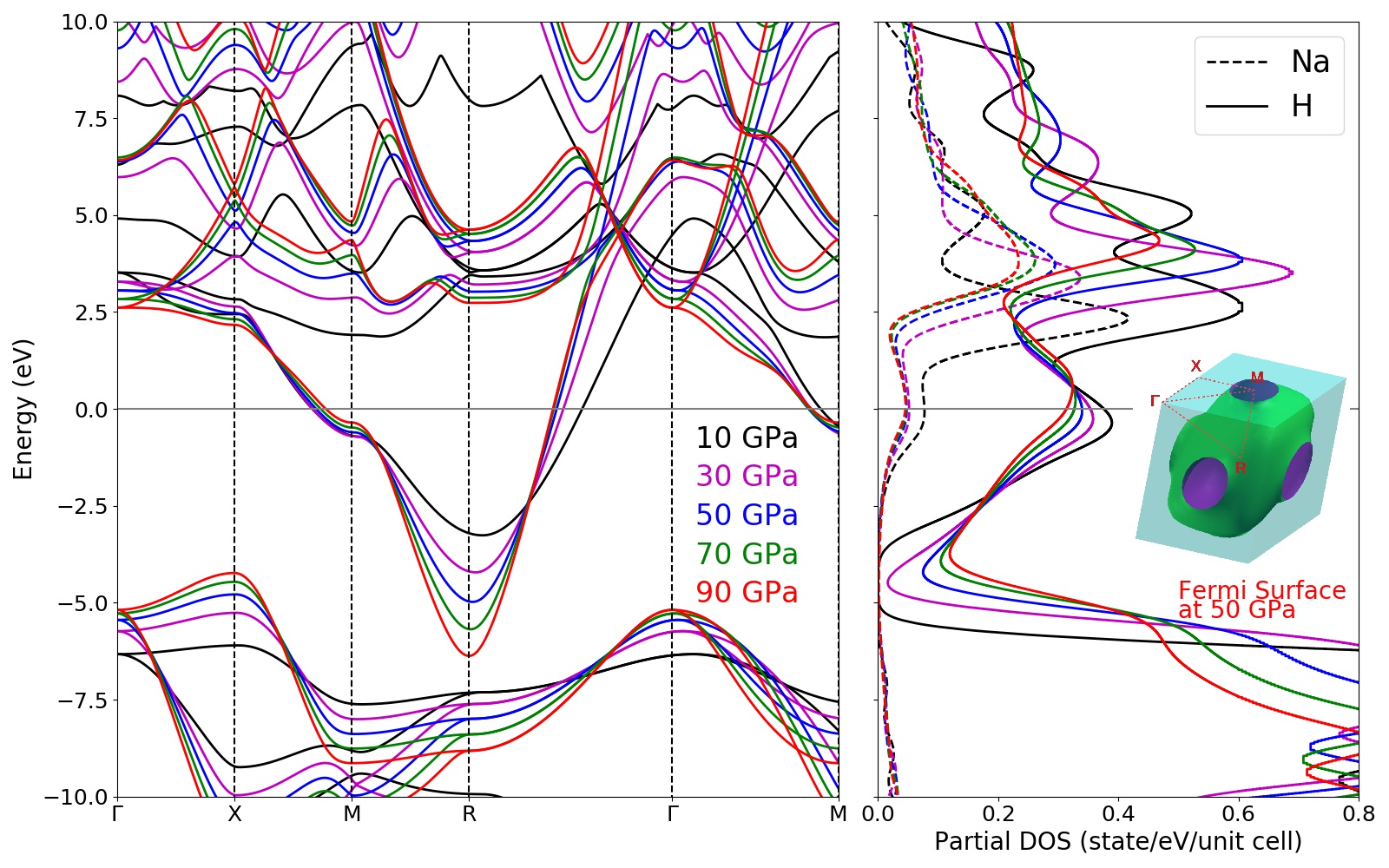}
\put (1,57) {\normalsize\textbf a}
\put (75,57) {\normalsize\textbf b}
\end{overpic}
\caption{
(a) Electronic band structures and (b) partial densities of states
projected onto Na and H atoms for $R\bar{3}m$ NaH$_6$ at 10, 30,
50, 70, and 90~GPa, shown in black, magenta, blue, green, and red,
respectively. The Fermi level is set to zero. The Fermi surface at
50~GPa is shown in the inset of (b).
}
\label{fig-ebandR-3m}
\end{figure}

The rapid increase of $T_\mathrm{c}$ of $R\bar{3}m$ NaH$_6$ with pressure
raises the question of what controls it. The two phases have comparable
densities and similar values of $N(E_\mathrm{F})$, $N_\mathrm{H}(E_\mathrm{F})$,
and $N_\mathrm{Na}(E_\mathrm{F})$, and the electronic band structure of
$R\bar{3}m$ NaH$_6$ near $E_\mathrm{F}$ changes only slightly with pressure
(Fig.~\ref{fig-ebandR-3m}). These observations suggest that the unusual
pressure dependence of $T_\mathrm{c}$ is primarily associated with changes in
the hydrogen sublattice rather than with major changes in the electronic
structure.

Figure~\ref{fig-pressureR-3m} suggests two structural parameters that
may be particularly relevant. First, the intermolecular H--H distance
$d_\mathrm{i}$ in $R\bar{3}m$ NaH$_6$ is considerably larger than that in
$Pm\bar{3}m$ NaH$_6$ at low pressures and decreases rapidly upon
compression. Second, $\theta$, the angle between neighboring
H$_2$-like units, is close to zero at low pressure and increases
rapidly with compression. To determine how these structural changes
affect the EPC, we examine the mode-resolved phonon linewidths.

As shown in Figs.~\ref{fig-a2F} and \ref{fig-a2FR-3m}, the phonon
linewidths of both $Pm\bar{3}m$ and $R\bar{3}m$ NaH$_6$ are largest
near the $\Gamma$ point. We therefore analyze the mode-specific phonon
linewidth $\gamma$ at $\Gamma$. As shown in
Table~\ref{table-omegagamma}, 12 of the 21 phonon modes have zero
$\gamma$ for both phases at 50~GPa; we refer to these modes as EPC
inactive, and to the remaining nine modes, which have nonzero $\gamma$,
as EPC active. Remarkably, all 12 EPC-inactive modes are non-Raman
modes, being either infrared active (I) or silent (S), whereas every
Raman-active (R) mode is EPC active with a large linewidth
($\gamma\geq 500$~GHz). The only EPC-active modes that are not Raman
active are the silent $A_\mathrm{2g}$ mode of $R\bar{3}m$ and the
silent $T_\mathrm{1g}$ modes of $Pm\bar{3}m$; their linewidths (15.4 and
3.0~GHz, respectively) are more than an order of magnitude smaller
than those of any Raman-active mode.

To determine whether this correlation extends beyond NaH$_6$, we
examined the EPC and Raman activities of phonon modes from 102
$X_m$H$_n$ hydride superconductors (X = Li, Ne, Na, Mg, Al, Si, P,
S, Cl, K, Ca, and Sc) at 200~GPa. For the 1,215 modes classified as
Raman active, the mean and median values of $\gamma$ are 1,356 and
472~GHz, respectively. In contrast, the corresponding values for the
1,110 non-Raman modes (infrared active or silent) are only 32 and
0.6~GHz, respectively. These results reveal a strong statistical
correlation between Raman activity and EPC strength in this set of
hydride superconductors.

\begin{table}[htbp]
\caption{
Mode symmetry, activity, phonon frequency $\omega$, and linewidth
$\gamma$ of the $\Gamma$-point phonons of $R\bar{3}m$ and $Pm\bar{3}m$
NaH$_6$ at 50~GPa. The activity is infrared (I), Raman (R), or silent
(S), i.e., neither infrared nor Raman active.
}
\label{table-omegagamma}
\centering
\begin{tabular}{|c|c|c|c|c|c|c|}
\hline
Phase & Mode \# & Label & Symm & Activity & $\omega$ (THz) & $\gamma$ (GHz) \\
\hline
 \multirow{14}{*}{$R\bar{3}m$}
&1 & & A$_\mathrm{2u}$ & I & 0.0 & 0.0 \\
&2-3 & & E$_\mathrm{u}$ & I & 0.0 & 0.0 \\
&4 & & A$_\mathrm{2u}$ & I & 12.7 & 0.0 \\
&5-6 & & E$_\mathrm{u}$ & I & 18.0 & 0.0 \\
&7 & A1 & A$_\mathrm{1g}$ & R & 25.7 & 519.8 \\
&8-9 & & E$_\mathrm{u}$ & I & 28.4 & 0.0 \\
&10-11 & E1 & E$_\mathrm{g}$ & R & 28.7 & 1033.2 \\
&12 & & A$_\mathrm{1u}$ & S & 30.1 & 0.0 \\
&13 & & A$_\mathrm{2u}$ & I & 34.6 & 0.0 \\
&14-15 & E2 & E$_\mathrm{g}$ & R & 37.5 & 999.8 \\
&16-17 & & E$_\mathrm{u}$ & I & 38.4 & 0.0 \\
&18 & A2 & A$_\mathrm{2g}$ & S & 44.2 & 15.4 \\
&19-20 & E3 & E$_\mathrm{g}$ & R & 86.4 & 2488.6 \\
&21 & A3 & A$_\mathrm{1g}$ & R & 97.2 & 1107.1 \\
\hline
 \multirow{8}{*}{$Pm\bar{3}m$}
&1-3 & & T$_\mathrm{1u}$ & I & 0.0 & 0.0 \\
&4-6 & & T$_\mathrm{1u}$ & I & 15.1 & 0.0 \\
&7-9 & T1 & T$_\mathrm{2g}$ & R & 25.8 & 3660.5 \\
&10-12 & & T$_\mathrm{2u}$ & S & 26.3 & 0.0 \\
&13-15 & & T$_\mathrm{1u}$ & I & 32.8 & 0.0 \\
&16-18 & T2 & T$_\mathrm{1g}$ & S & 40.3 & 3.0 \\
&19-20 & E3 & E$_\mathrm{g}$ & R & 76.7 & 2182.9 \\
&21 & A3 & A$_\mathrm{1g}$ & R & 93.0 & 873.5 \\
\hline
\end{tabular}
\end{table}

\begin{figure}[htp]
\centering
\begin{overpic}[width=0.48\textwidth]{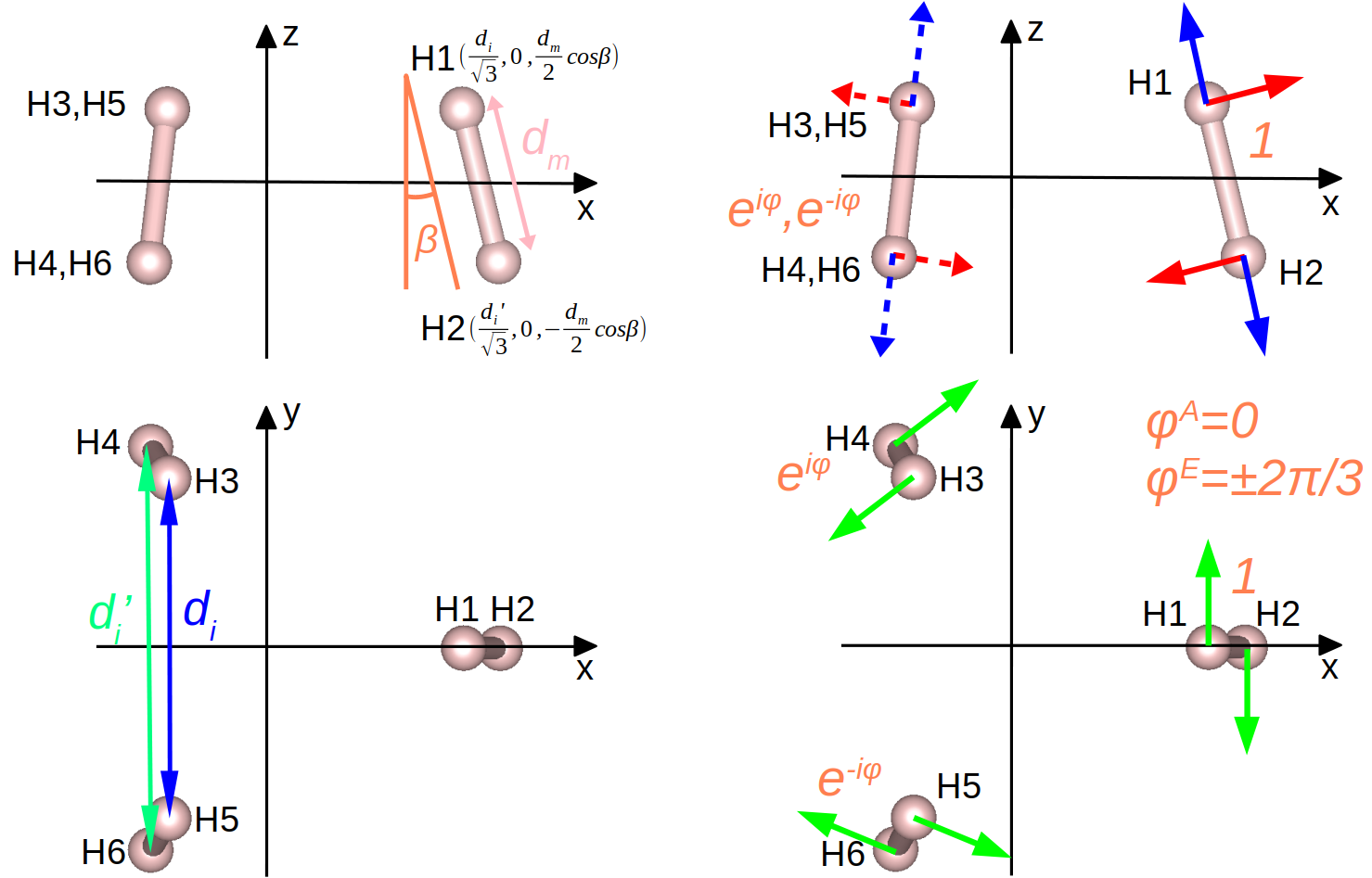}
\put (0,60) {\normalsize\textbf a}
\put (0,28) {\normalsize\textbf b}
\put (52,60) {\normalsize\textbf c}
\put (52,28) {\normalsize\textbf d}
\end{overpic}
\caption{
(a), (b) Coordinates and characteristic distances of neighboring
H$_2$-like units in $R\bar{3}m$ NaH$_6$ at 50~GPa.
(c), (d) Schematic illustration of the EPC-active phonon modes.
Type 1, Type 2, and Type 3 modes are indicated by red, green, and blue arrows, respectively, with the corresponding phase factors
shown in orange. The dashed arrows in (c) indicate displacements
that are not parallel to the $x$--$z$ plane. The phase factors of
the H1--H2, H3--H4, and H5--H6 units are $1$, $\mathrm{e}^{\mathrm{i}\varphi}$,
and $\mathrm{e}^{-\mathrm{i}\varphi}$, respectively, where $\varphi=0$ for the
in-phase A1--A3 modes and $\varphi=\pm2\pi/3$ for the out-of-phase
E1--E3 modes.
}
\label{fig-9modes}
\end{figure}

\begin{figure}[htp]
\centering
\begin{overpic}[width=0.48\textwidth]{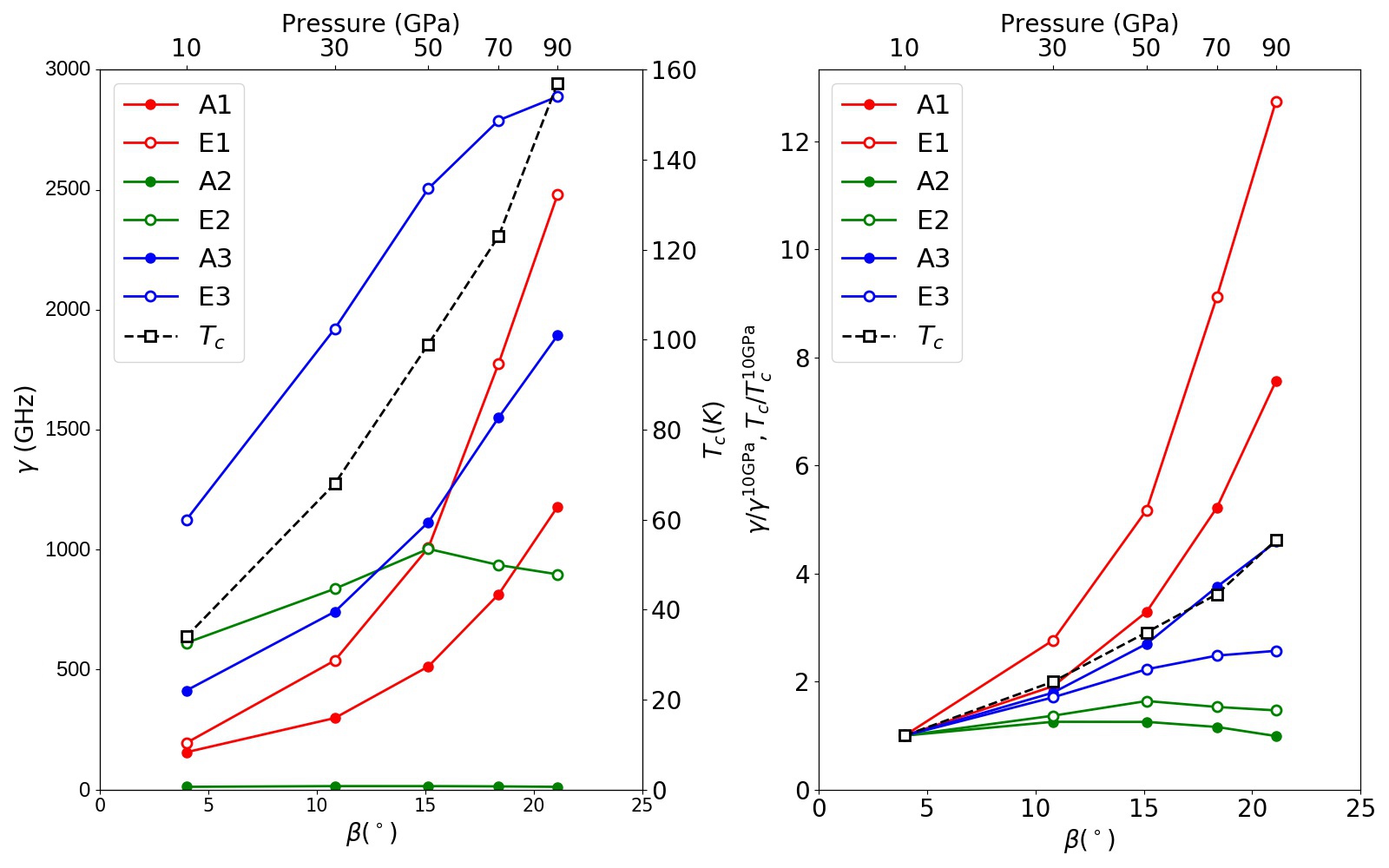}
\put (0,58) {\normalsize\textbf a}
\put (53,58) {\normalsize\textbf b}
\end{overpic}
\caption{
(a) Calculated $T_\mathrm{c}$ and mode-specific $\Gamma$-point phonon
linewidths $\gamma$ of $R\bar{3}m$ NaH$_6$ as functions of $\beta$ (bottom
axis) and pressure (top axis), where $\beta$ is the angle between the H1–-H2 axis and the $z$ axis. (b) $T_\mathrm{c}$ and $\gamma$ normalized
to their values at 10~GPa.
}
\label{fig-Gamma}
\end{figure}

The mode-resolved linewidths of $R\bar{3}m$ NaH$_6$ depend strongly on the
tilt $\beta$ of the H$_2$-like units, defined as the angle between the H1--H2
axis and the threefold axis [Fig.~\ref{fig-9modes}]. $\beta$ is related to the angle $\theta$ between neighboring units by
$\sin\beta=(2/\sqrt{3})\sin(\theta/2)$. Figure~\ref{fig-Gamma}
shows that, for $\beta$ between $4^\circ$ (10~GPa) and $21^\circ$ (90~GPa), the
linewidths of A1, E1, A3, and E3 increase monotonically with $\beta$, whereas
that of E2 increases up to $\beta\approx15^\circ$ (50~GPa) and decreases
slightly at larger $\beta$. The linewidth of the Raman-silent mode A2 remains
more than an order of magnitude smaller than those of the Raman-active modes.

A vibrational mode is Raman active when the electronic polarizability
changes in response to the vibration. Because the polarizability is a
symmetric second-rank tensor, only modes that transform as irreducible
representations contained in such a tensor ($A_\mathrm{1g}$ and
$E_\mathrm{g}$ in $D_\mathrm{3d}$; $A_\mathrm{1g}$, $E_\mathrm{g}$, and
$T_\mathrm{2g}$ in $O_\mathrm{h}$) can change it to first order in the
atomic displacements. The strong correlation between Raman activity and
EPC activity therefore suggests that the symmetry properties governing
the change in polarizability may also provide insight into the coupling
between electronic states and lattice vibrations.

To explore this idea, we use a minimal geometrical model. It is a
geometrical proxy: the tensor $\bm D$ defined below transforms like the
polarizability, but it is neither the electronic polarizability nor an
electron--phonon matrix element. Because the Na contribution to the
coupling is less than 5\%, we consider only the hydrogen sublattice and
describe it by the second-moment tensor
\begin{equation}
\bm D=\sum_{\mathrm H}\bm r\,\bm r^{T},
\end{equation}
where $\bm r=(x,y,z)$ are the H coordinates measured from the center of
the three neighboring H$_2$-like units (the six H atoms are symmetry
equivalent, so they carry equal weights). $\bm D$ differs from the
traceless quadrupole tensor only by the isotropic term
$(\mathrm{tr}\,\bm D/3)\,\mathbf{I}$, which affects only the totally
symmetric ($A_\mathrm{1g}$) channel; we keep it.

With the $z$ axis along the threefold axis, the H1--H2 unit lies in the
$x$--$z$ plane,
\begin{align}
\bm r_{\mathrm{H1}}&=\left(\frac{d_\mathrm{i}}{\sqrt{3}},\,0,\,\frac{d_\mathrm{m}\cos\beta}{2}\right),\nonumber\\
\bm r_{\mathrm{H2}}&=\left(\frac{d_\mathrm{i}'}{\sqrt{3}},\,0,\,-\frac{d_\mathrm{m}\cos\beta}{2}\right),
\end{align}
where $d_\mathrm{i}'-d_\mathrm{i}=\sqrt{3}\,d_\mathrm{m}\sin\beta$. The unit vector along H1--H2,
the in-plane unit vector perpendicular to it, and the $y$ axis are
$\hat{\bm a}=(-\sin\beta,0,\cos\beta)$,
$\hat{\bm p}=(\cos\beta,0,\sin\beta)$, and
$\hat{\bm y}=(0,1,0)$, respectively.

The nine EPC-active modes correspond to three types of relative motion of
the two H atoms of each unit [Fig.~\ref{fig-9modes}]: Type~1 (A1, E1) is a
libration in the $x$--$z$ plane, $\bm u=\hat{\bm p}$; Type~2 (A2, E2) is a
libration out of this plane, $\bm u=\hat{\bm y}$; and Type~3 (A3, E3) is
the stretching of the H$_2$-like unit, $\bm u=\hat{\bm a}$. In all these
modes the two H atoms of a unit move oppositely, $\bm r_{\mathrm{H1}}\to
\bm r_{\mathrm{H1}}+\delta\bm u$ and $\bm r_{\mathrm{H2}}\to
\bm r_{\mathrm{H2}}-\delta\bm u$, and the three units move with phase
factors $1$, $\mathrm{e}^{\mathrm{i}\varphi}$, $\mathrm{e}^{-\mathrm{i}\varphi}$, with $\varphi=0$ for the
A modes and $\varphi=\pm2\pi/3$ for the E modes (real partners of a
degenerate pair are obtained from amplitudes $\cos(2\pi j/3+\chi)$ on unit
$j$). To first order in $\delta$,
\begin{align}
\Delta\bm D&=\sum_{j=0}^{2}\mathrm{e}^{\mathrm{i}j\varphi}\,\bm R^{j}\,\Delta\bm D_0\,\bm R^{-j},\nonumber\\
\Delta\bm D_0&=d_\mathrm{m}\,\delta\,\bigl(\hat{\bm a}\,\bm u^{T}+\bm u\,\hat{\bm a}^{T}\bigr),
\end{align}
where $\bm R$ is the $120^\circ$ rotation about $z$ and
$\Delta\bm D_0$ follows from $\bm r_{\mathrm{H1}}-\bm r_{\mathrm{H2}}=d_\mathrm{m}\hat{\bm a}$.
The intermolecular distances $d_\mathrm{i}$ and $d_\mathrm{i}'$ cancel.
Within this minimal geometric construction, $\Delta D$ vanishes to first
order for the infrared-active and silent modes. For A2, $\Delta\bm D_0\neq0$ but has only $xy$ and $yz$
components, which are $E_\mathrm{g}$-type and cancel in the in-phase
threefold sum.

The threefold sum projects $\Delta\bm D$ onto the symmetry channels of
$D_\mathrm{3d}$. For the A modes, $\Delta\bm D=\mathrm{diag}(P,P,Z)$ has an
in-plane amplitude $P$ ($xx=yy$) and an amplitude $Z$ along the threefold
axis ($zz$). For the E modes, $\Delta\bm D$ is traceless and has two doublet
components, an in-plane one, $w=(\Delta D_{xx}-\Delta D_{yy})/2+\mathrm{i}\,\Delta D_{xy}$, and one that
involves the threefold axis, $v=\Delta D_{xz}+\mathrm{i}\,\Delta D_{yz}$. For the real partner
in which the H1--H2 unit carries the full amplitude, $\Delta\bm D$ equals
$\left(\begin{smallmatrix}A&0&B\\0&-A&0\\B&0&0\end{smallmatrix}\right)$ (Types 1 and 3) or
$\left(\begin{smallmatrix}0&A&0\\A&0&-B\\0&-B&0\end{smallmatrix}\right)$ (Type 2). Other real
combinations of the two partners multiply $w$ and $v$ by opposite phase
factors, $\mathrm{e}^{\mp \mathrm{i}\chi}$, so
$|w|=|A|$, $|v|=|B|$, and $\mathrm{Re}(wv)=AB$ for every combination;
these are the quantities that enter below. The amplitudes (in units of
$d_\mathrm{m}\delta$) are listed in Table~\ref{table-dD}. We have verified
numerically that Table~\ref{table-dD} follows from the equations above.
As a check, at $\beta=\arccos(1/\sqrt{3})\approx54.7^\circ$
($\theta=90^\circ$, the $Pm\bar{3}m$ geometry) these tensors reduce to
cubic irreducible components: A3 is isotropic ($A_\mathrm{1g}$), E3 is
$E_\mathrm{g}$, and A1, E1, and E2 are $T_\mathrm{2g}$, as expected for
the stretching and librational modes of $Pm\bar{3}m$ NaH$_6$
(Table~\ref{table-omegagamma}).

\begin{table}[htbp]
\caption{
Amplitudes of $\Delta\bm D$ for the EPC-active modes of $R\bar{3}m$ NaH$_6$,
in units of $d_\mathrm{m}\delta$. $P$ and $A$ are in-plane components
($xx$, $yy$, $xy$); $Z$ and $B$ involve the threefold axis ($zz$, $xz$,
$yz$). For E modes, the values refer to the real partner in which the H1--H2 unit
carries the full amplitude. The first-order $\Delta\bm D$ of A2 vanishes.
}
\label{table-dD}
\centering
\begin{tabular}{|c|c|c||c|c|c|}
\hline
Mode & $P$ & $Z$ & Mode & $A$ & $B$ \\
\hline
A1 & $-\tfrac{3}{2}\sin2\beta$ & $3\sin2\beta$ & E1 & $-\tfrac{3}{4}\sin2\beta$ & $\tfrac{3}{2}\cos2\beta$ \\
A2 & 0 & 0 & E2 & $-\tfrac{3}{2}\sin\beta$ & $-\tfrac{3}{2}\cos\beta$ \\
A3 & $3\sin^2\beta$ & $6\cos^2\beta$ & E3 & $\tfrac{3}{2}\sin^2\beta$ & $-\tfrac{3}{2}\sin2\beta$ \\
\hline
\end{tabular}
\end{table}

To relate $\Delta\bm D$ to the linewidth we make a model assumption: the
electron--phonon coupling is, to leading order, a linear function of the
components of $\Delta\bm D$, so that $\gamma$ is a quadratic form in them, with
coefficients that are independent of $\beta$ over the range considered.
Symmetry constrains this form but does not fix its coefficients. The form
must be invariant under $D_\mathrm{3d}$. Components that belong to different
irreducible representations cannot be combined into an invariant
quadratic form, whereas two copies of the same representation can. The
$A_\mathrm{1g}$ channel contains two components, $P$ and $Z$, which may
mix; the $E_\mathrm{g}$ channel contains two copies of the doublet, $w$ and
$v$, which may also mix (the mirror symmetry excludes
$\mathrm{Im}(wv)$). Therefore 
\begin{align}
\gamma^{\mathrm{A}}
&\propto k_P P^2+2k_{PZ}PZ+k_Z Z^2,
\label{eq:gamma-A}\\
\gamma^{\mathrm{E}}
&\propto k_A |w|^2+2k_{AB}\,\mathrm{Re}(wv)+k_B |v|^2
\nonumber\\
&= k_A A^2+2k_{AB}AB+k_B B^2,
\label{eq:gamma-E}
\end{align}
where the constants $k$ are set by the electronic structure, and the
mode-normalization factors are absorbed into them. The A and E channels
carry independent constants. This quadratic form is a hypothesis of
the model, not a consequence of symmetry.

The A2 linewidth is consistent with the absence of a first-order $\Delta\bm D$;
its small but nonzero value shows that $\Delta\bm D$ does not capture all of the
coupling. In Table~\ref{table-dD}, the magnitudes of the in-plane amplitudes
$P$ and $A$ increase with $\beta$ for all five modes in the range of
Fig.~\ref{fig-Gamma}, whereas the magnitudes of the axial amplitudes $Z$ (A3)
and $B$ (E1, E2) decrease. If the axial amplitudes dominated, the linewidths
of A3, E1, and E2 would therefore decrease with increasing $\beta$, contrary
to the DFT results. Within this phenomenological model, the increasing
trends are thus consistent with a significant contribution from the in-plane
components of $\Delta\bm D$, i.e., from the response of the hydrogen sublattice
perpendicular to the threefold axis. This is a qualitative interpretation of
the calculated trends: it depends on the coefficients $k$, which are not determined here; it is
not a symmetry result, and it does not account for the magnitudes. In
particular, the linewidths are already substantial at $\beta\approx4^\circ$,
and the model, which uses pure displacement types, neglects the mixing of
E1--E3 allowed by symmetry and any contribution beyond the tensor
$\Delta\bm D$. The increase of $\beta$ with pressure is consistent with the pressure-induced
increase of the linewidths of the Raman-active modes, and hence of
$\alpha^2F(\omega)$ near their frequencies (librational modes at 26--38~THz
and stretching modes at 86--97~THz; Table~\ref{table-omegagamma}), and with the
increase of $\lambda$ and $T_\mathrm{c}$ in $R\bar{3}m$ NaH$_6$, but it does
not by itself establish the mechanism.

\section{\label{sec:conclusion}Conclusion}

In summary, we have presented a comprehensive first-principles study of
the structural, vibrational, and superconducting properties of
NaH$_6$ over a broad pressure range. The cubic $Pm\bar{3}m$ phase is
dynamically stable down to 50~GPa and exhibits remarkably high
superconducting transition temperatures, with $T_\mathrm{c}=230$~K at 50~GPa
and 215~K at 200~GPa. At lower pressures, the rhombohedral
$R\bar{3}m$ phase becomes energetically favorable and remains
dynamically stable down to 10~GPa, where its calculated $T_\mathrm{c}$ is 34~K.
In contrast to the weak pressure dependence of $T_\mathrm{c}$ in
the $Pm\bar{3}m$ phase, $T_\mathrm{c}$ of $R\bar{3}m$ NaH$_6$ increases
rapidly with compression, reaching 99~K at 50~GPa and 157~K at
90~GPa. The calculated enthalpies indicate that the $R\bar{3}m$ and
$Pm\bar{3}m$ phases become energetically competitive near 114~GPa.
Finite-temperature molecular-dynamics simulations and phonon
renormalization calculations further indicate that the relevant
structures are robust at finite temperature and that anharmonic
effects on the lattice dynamics are relatively small.

A particularly unusual feature of NaH$_6$ is that the hydrogen atoms
in both the $Pm\bar{3}m$ and $R\bar{3}m$ phases form distinct
H$_2$-like units with H--H bonds substantially elongated relative to
those in an isolated H$_2$ molecule. Hydrogen-rich structures
containing such molecular-like units have generally been considered
less favorable for high-$T_\mathrm{c}$ superconductivity because strong H--H
bonding tends to place the corresponding bonding states well below
the Fermi level \cite{sun2019route}. NaH$_6$ demonstrates that the
presence of H$_2$-like units does not necessarily preclude strong
electron--phonon coupling or high-temperature superconductivity.
Instead, our results indicate that the structural arrangement and
orientational degrees of freedom of these units play an important
role in determining the superconducting properties.

Our mode-resolved analysis further reveals a strong correlation
between Raman activity and electron--phonon coupling in NaH$_6$ and
across a broader set of hydride superconductors. In $R\bar{3}m$ NaH$_6$, the
$\Gamma$-point linewidths of the Raman-active modes increase with the tilt
$\beta$ of the H$_2$-like units over the calculated pressure range
10--90~GPa (that of E2 up to about 50~GPa), whereas the Raman-silent mode A2
remains nearly zero. To interpret these trends, we introduced a minimal
geometrical model based on the vibration-induced change of the
second-moment tensor $\bm D$ of the hydrogen sublattice, which has
the symmetry of the polarizability but is not itself an electron--phonon
matrix element. Within this model, the calculated trends are consistent with
a significant contribution from the components of $\Delta\bm D$
perpendicular to the threefold axis. The model is qualitative and does not
account for the magnitudes of the linewidths, but together with the growth of
$\alpha^2F$ in the 20--40~THz range where the librational modes are located,
it suggests that the pressure-induced tilt of the H$_2$-like units is an
important factor in the rapid increase of $T_\mathrm{c}$ in $R\bar{3}m$
NaH$_6$. These results suggest that the geometry and orientational dynamics
of H$_2$-like units may provide an additional degree of freedom for
controlling electron--phonon coupling in hydrogen-rich superconductors, and
motivate calculations based on the full phonon eigenvectors and further
investigation of this mechanism in other molecular-like hydrides.

\begin{acknowledgments}
T.C. acknowledges support from the NIST Fellowship Program during the initial stages of this work.
\end{acknowledgments}

\appendix

\section{\label{sec:computational}Computational Details}

Variable-composition searches of the binary sodium hydride
Na$_m$H$_n$ system at 50, 60, 70, 100, 150, and 200~GPa were
performed using the USPEX evolutionary algorithm. Structural
relaxations were performed using the VASP package with the
Perdew--Burke--Ernzerhof (PBE) exchange-correlation functional and
the projector-augmented-wave (PAW) method \cite{paier2005perdew,perdew1996generalized,blochl1994projector}.
A plane-wave kinetic-energy cutoff of 650~eV was used, and the
Brillouin zone was sampled using $\Gamma$-centered k-point meshes \cite{monkhorst1976special}
with a resolution of $0.04 \times 2\pi$ \AA$^{-1}$.

Density-functional theory calculations of the electronic structure,
phonons, and electron--phonon coupling were performed using
QUANTUM ESPRESSO with the PBE exchange-correlation functional and scalar-relativistic ultrasoft pseudopotentials
\cite{vanderbilt1990soft}. Kinetic-energy cutoffs of 120~Ry and
1,200~Ry were used for the wave functions and charge density,
respectively. A $32\times32\times32$ k-point mesh was used for
structural optimization, with the atomic positions relaxed until all
residual force components were smaller than $10^{-6}$~Ry/a.u.
Electronic band structures and densities of states were calculated
using a $64\times64\times64$ k-point mesh. For the phonon and EPC
calculations, $16\times16\times16$ q-point and
$16\times16\times16$ k-point meshes were used for $Pm\bar{3}m$
NaH$_6$, while $8\times8\times8$ q-point and
$16\times16\times16$ k-point meshes were used for $R\bar{3}m$
NaH$_6$.

Molecular-dynamics (MD) simulations were performed using VASP with
the PBE functional and PAW method. A $5\times5\times5$ supercell
containing 875 atoms was used. Brillouin-zone sampling employed a
$2\times2\times2$ k-point mesh for $Pm\bar{3}m$ NaH$_6$ at
50~GPa and a $3\times3\times3$ k-point mesh for $Pm\bar{3}m$
NaH$_6$ at 200~GPa and $R\bar{3}m$ NaH$_6$ at 110~GPa. The time
step was 0.5~fs. The simulations consisted of 15,700 steps for
$Pm\bar{3}m$ NaH$_6$ at 50~GPa and 7,600 steps for
$Pm\bar{3}m$ NaH$_6$ at 200~GPa and $R\bar{3}m$ NaH$_6$ at
110~GPa.

Renormalized phonon frequencies were obtained using DynaPhoPy
\cite{carreras2017dynaphopy}. Harmonic phonon frequencies were
calculated using the finite-displacement method with VASP and
Phonopy. The supercell sizes and k-point meshes were the same as
those used in the corresponding MD simulations.

\bibliographystyle{apsrev4-1}
\bibliography{mybib}

\clearpage

\setcounter{figure}{0}
\renewcommand{\thefigure}{S\arabic{figure}}
\setcounter{table}{0}
\renewcommand{\thetable}{S\arabic{table}}
\setcounter{equation}{0}
\renewcommand{\theequation}{S\arabic{equation}}

\onecolumngrid
\input{Supplementary}

\end{document}

%% file: Supplementary.tex
\begin{center}
    \LARGE \textbf{Supplementary Material} \\[0.5cm]
    \large \textbf{H$_2$-Like Units and Pressure-Tunable High-$T_\mathrm{c}$ Superconductivity in NaH$_6$} \\[0.4cm]
    \normalsize Tianran Chen$^1$ and Taner Yildirim$^2$ \\[0.2cm]
    \small \textit{$^{1}$Department of Physics and Astronomy, University of Tennessee, Knoxville, Tennessee 37996, USA \\
    $^{2}$NIST Center for Neutron Research, National Institute of Standards and Technology, Gaithersburg, MD 20899-8562, USA} \\[1cm]
\end{center}

\FloatBarrier

\subsection*{S1. Structure searches in Na--H and related binary hydrides}

\begin{figure}[!htbp]
\centering
\begin{overpic}[width=0.9\columnwidth]{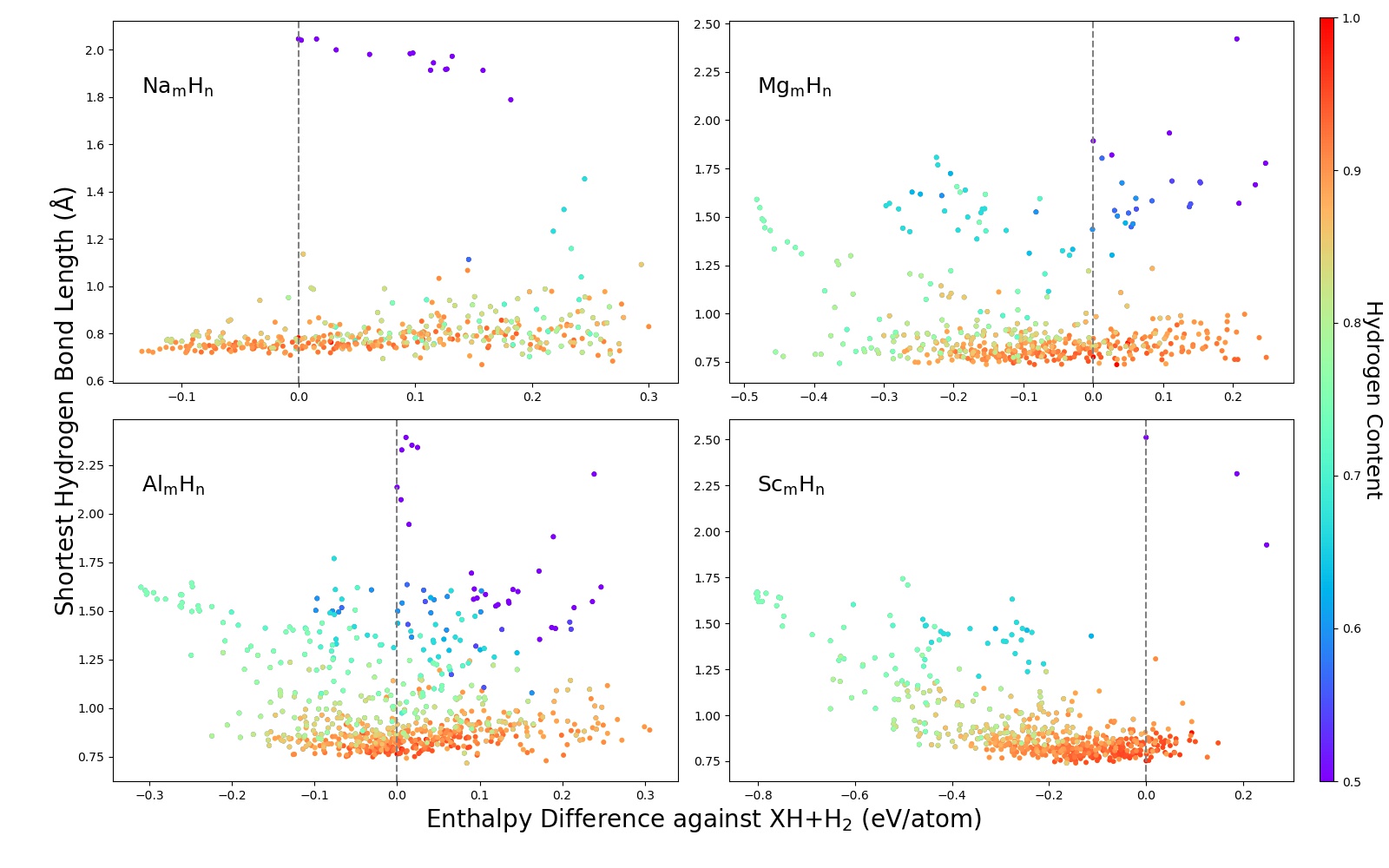}
\put (9,58) {\large\textbf a} 
\put (54,58) {\large\textbf b} 
\put (9,29) {\large\textbf c} 
\put (54,29) {\large\textbf d} 
\end{overpic}
\caption{Formation enthalpies and shortest H--H distances of $X_m\mathrm{H}_n$ structures for (a)~$X = \mathrm{Na}$, (b)~$\mathrm{Mg}$, (c)~$\mathrm{Al}$, and (d)~$\mathrm{Sc}$ at 200~GPa. Enthalpies are referenced to $m\,X\mathrm{H}+\frac{n-m}{2}\,\mathrm{H}_2$. Colors denote the hydrogen fraction.}
\label{fig-hhbond}
\end{figure}

We performed USPEX evolutionary searches for Na$_m$H$_n$, Mg$_m$H$_n$, Al$_m$H$_n$, and Sc$_m$H$_n$ at 200~GPa. Figure~\ref{fig-hhbond} compares the formation enthalpies and shortest H--H distances of the resulting structures. Nearly all low-enthalpy sodium hydrides contain H$_2$-like units with shortest H--H distances below 1.0~\AA, except for several NaH structures, shown as purple points with hydrogen fraction 0.5. In contrast, many low-enthalpy Mg, Al, and Sc hydrides do not contain such H$_2$-like units.

\begin{figure}[!htbp]
\centering
\begin{overpic}[width=0.9\columnwidth]{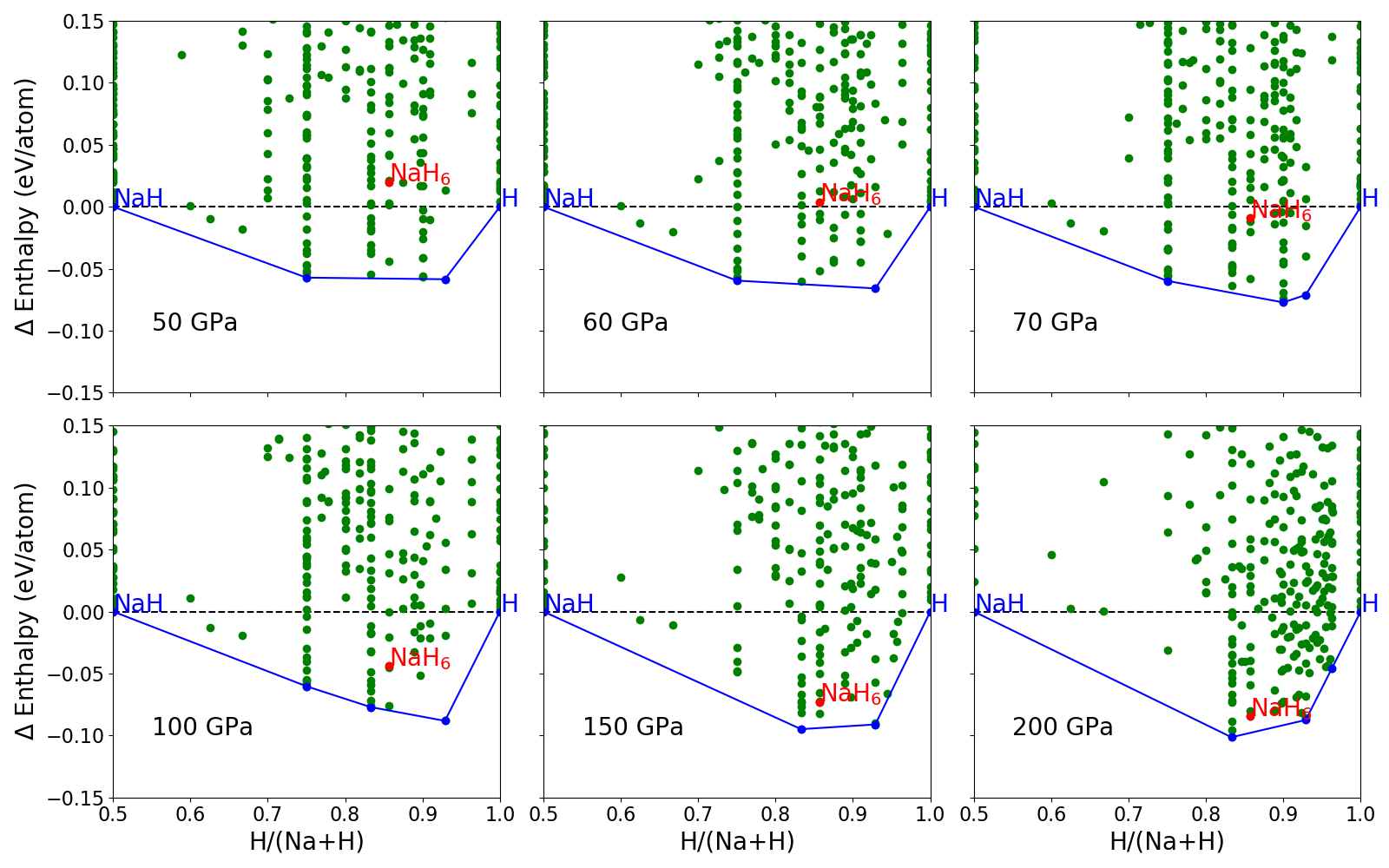}
\put (9,59) {\large\textbf a}
\put (40,59) {\large\textbf b}
\put (71,59) {\large\textbf c}
\put (9,29) {\large\textbf d}
\put (40,29) {\large\textbf e}
\put (71,29) {\large\textbf f}
\end{overpic}
\caption{Convex-hull diagrams of Na$_m$H$_n$ at (a) 50, (b) 60, (c) 70, (d) 100, (e) 150, and (f) 200~GPa. $Pm\bar{3}m$ NaH$_6$ is indicated by red points. Points below the black dashed line are stable against decomposition into NaH and H$_2$, whereas points on the blue convex hull are stable against decomposition into all phases included in the diagram.}
\label{fig-convexhull}
\end{figure}

Additional USPEX searches of Na$_m$H$_n$ were performed at 50, 60, 70, 100, 150, and 200~GPa. Previously reported stable compositions, including Na$_2$H$_3$, Na$_2$H$_5$, and Na$_2$H$_9$ \cite{baettig2011pressure}, were supplied to USPEX as seed structures. Figure~\ref{fig-convexhull} shows the resulting convex-hull diagrams with respect to NaH and H$_2$. The $Pm\bar{3}m$ NaH$_6$ phase is marked by red points. Its enthalpy relative to $\mathrm{NaH}+2.5\,\mathrm{H}_2$ is 0.02, 0.00, $-0.01$, $-0.04$, $-0.07$, and $-0.09$~eV/atom at 50, 60, 70, 100, 150, and 200~GPa, respectively.

\subsection*{S2. Superconducting properties of $Pm\bar{3}m$ NaH$_6$}

\begin{figure}[!htbp]
\centering
\begin{overpic}[width=0.5\columnwidth]{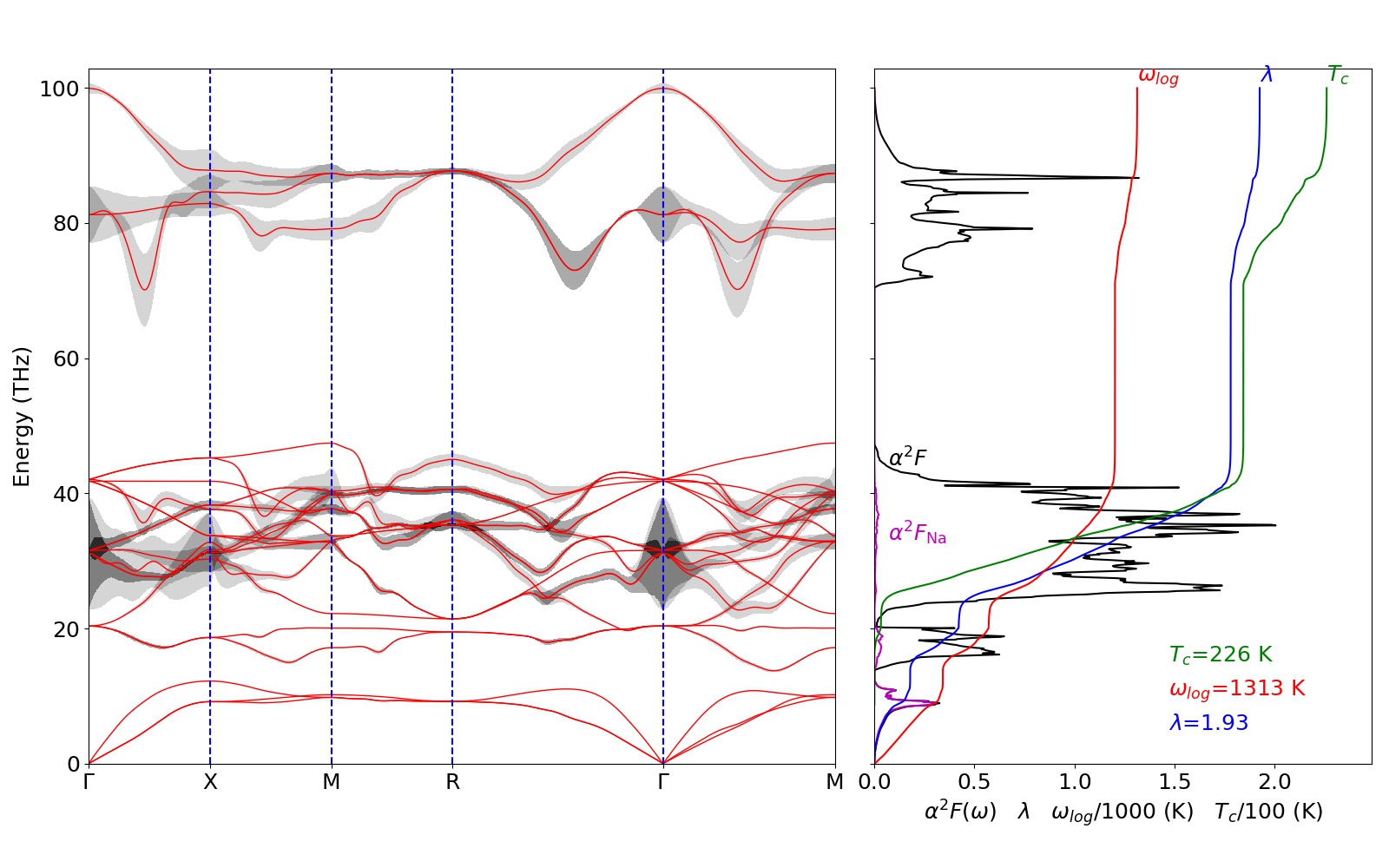}
\put (0,54) {\large\textbf a}
\put (65,54) {\large\textbf b}
\end{overpic}
\begin{overpic}[width=0.5\columnwidth]{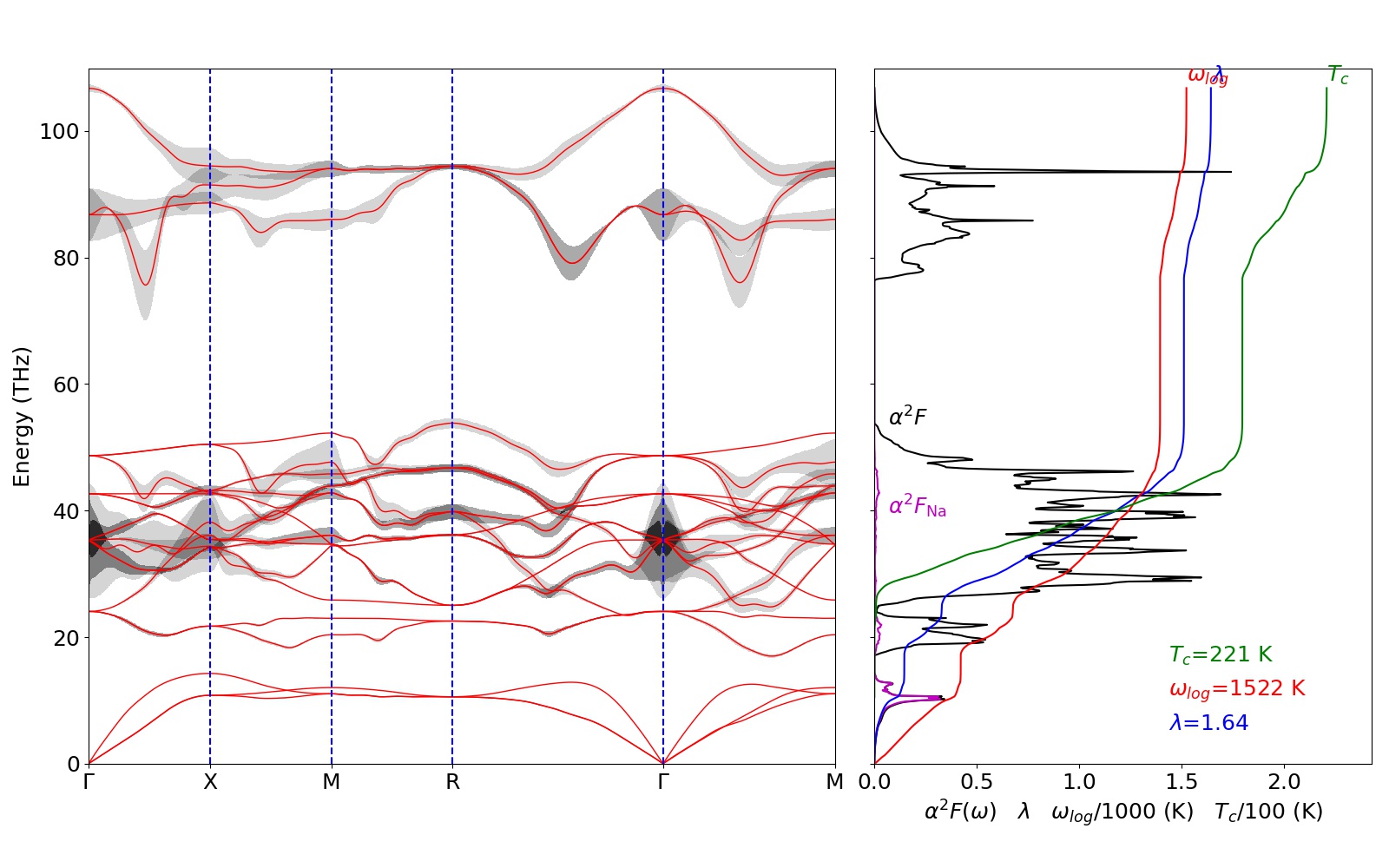}
\put (0,54) {\large\textbf c}
\put (65,54) {\large\textbf d}
\end{overpic}
\begin{overpic}[width=0.5\columnwidth]{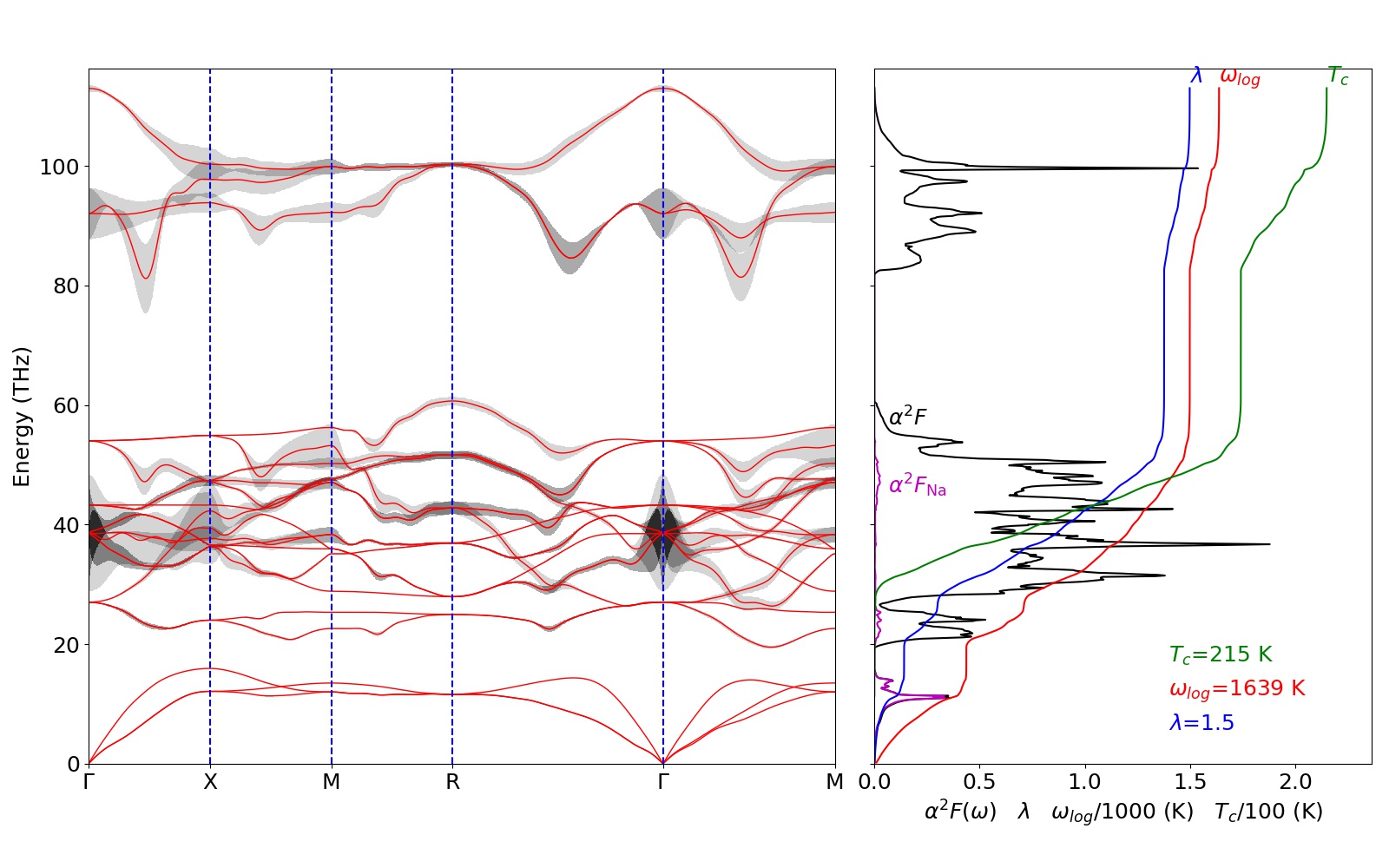}
\put (0,54) {\large\textbf e}
\put (65,54) {\large\textbf f}
\end{overpic}
\caption{(a), (c), and (e) Phonon dispersions of $Pm\bar{3}m$ NaH$_6$ at 100, 150, and 200~GPa, respectively. The gray shaded areas represent the phonon linewidths. (b), (d), and (f) Total and Na-projected Eliashberg spectral function $\alpha^2F(\omega)$, together with the integrated $T_\mathrm{c}$, $\omega_\mathrm{log}$, and $\lambda$. The black and purple curves show the total and Na-projected $\alpha^2F(\omega)$, respectively; the green, red, and blue curves show the integrated $T_\mathrm{c}$, $\omega_\mathrm{log}$, and $\lambda$, respectively.}
\label{fig-a2FPressure}
\end{figure}

The phonon dispersions, Eliashberg spectral functions $\alpha^2F(\omega)$, electron--phonon coupling (EPC) constants $\lambda$, logarithmic phonon frequencies $\omega_\mathrm{log}$, and transition temperatures $T_\mathrm{c}$ of $Pm\bar{3}m$ NaH$_6$ at 100, 150, and 200~GPa are shown in Fig.~\ref{fig-a2FPressure}. The corresponding superconducting parameters at all calculated pressures are listed in Table~\ref{table-Tc}.

\begin{table}[!htbp]
\caption{Calculated superconducting parameters, phonon gaps, and mode contributions for $Pm\bar{3}m$ NaH$_6$. $G_1$ is the gap between the acoustic and optical branches, and $G_2$ is the gap between the high-frequency H$_2$-like stretching band and the remaining optical modes. $P_\mathrm{Na}$ and $P_{\mathrm{H}_2}$ are the contributions of Na vibrations and of the high-frequency H$_2$-like stretching band, respectively, to $\alpha^2F(\omega)$.}
\label{table-Tc}
\centering
\begin{tabular}{|c|c|c|c|c|c|c|c|}
\hline
$P$ & $T_\mathrm{c}$ & $\omega_\mathrm{log}$ & $\lambda$ & $G_1$ & $G_2$ & $P_\mathrm{Na}$ & $P_{\mathrm{H}_2}$ \\
(GPa) & (K) & (K) &  & (THz) & (THz) & (\%) & (\%) \\
\hline
50  & 230 & 873  & 2.80 & 0.00 & 22.92 & 4.11 & 18.44 \\
100 & 226 & 1313 & 1.93 & 1.51 & 22.67 & 2.96 & 19.38 \\
150 & 221 & 1522 & 1.64 & 2.73 & 21.82 & 3.00 & 19.48 \\
200 & 215 & 1639 & 1.50 & 3.55 & 20.49 & 3.24 & 19.45 \\
\hline
\end{tabular}
\end{table}

\subsection*{S3. Thermal properties of $Pm\bar{3}m$ NaH$_6$}

\begin{figure}[!htbp]
\centering
\begin{overpic}[width=0.9\columnwidth]{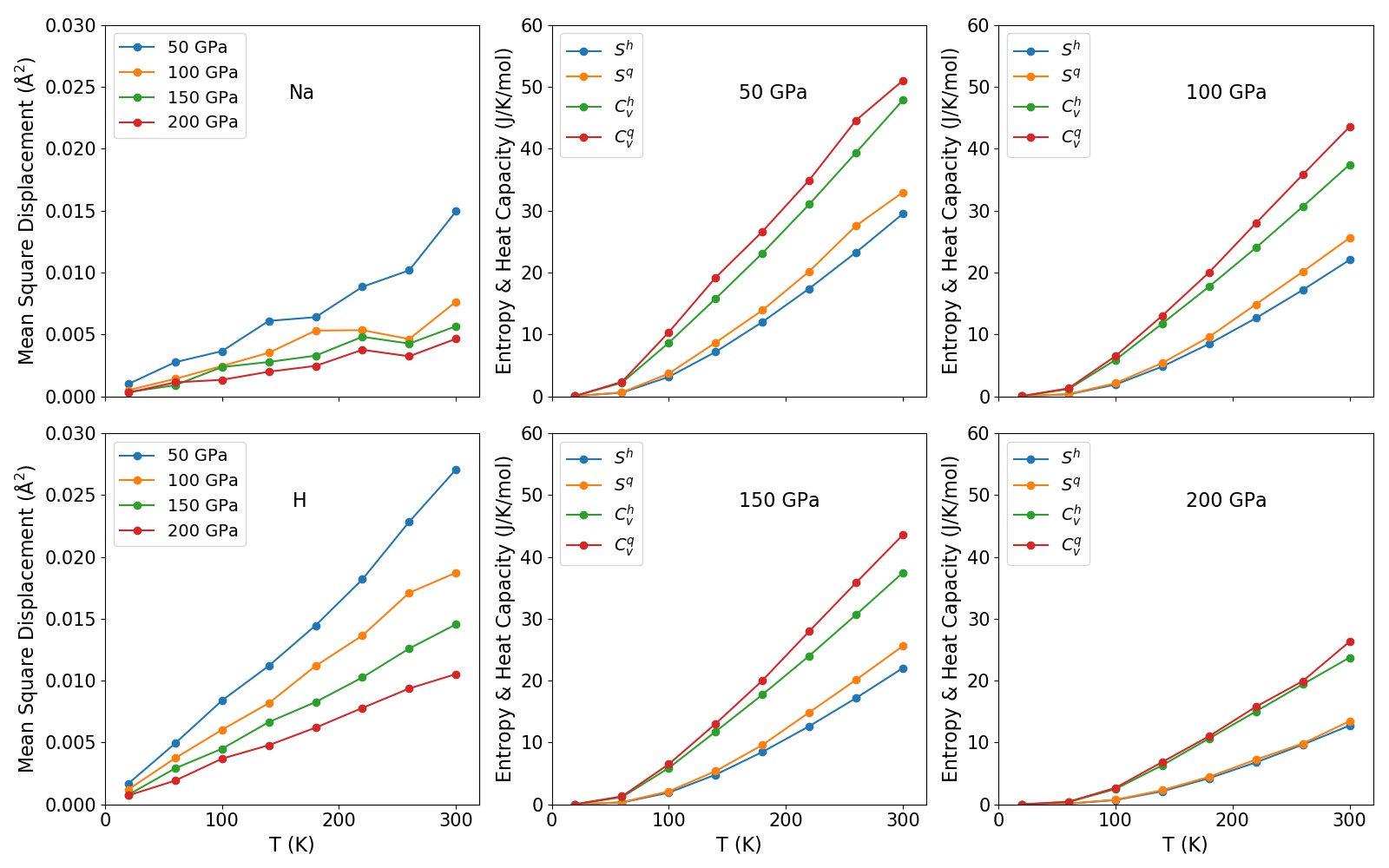}
\put (30,58) {\large\textbf a} 
\put (62,58) {\large\textbf b} 
\put (94,58) {\large\textbf c} 
\put (30,28) {\large\textbf d} 
\put (62,28) {\large\textbf e} 
\put (94,28) {\large\textbf f} 
\end{overpic}
\caption{Thermal properties from $2\times2\times2$ supercell molecular-dynamics simulations. (a), (d) Mean-square displacements of Na and H atoms as functions of temperature and pressure. (b), (c), (e), (f) Entropy and heat capacity obtained from harmonic ($S^\mathrm{h}$, $C_V^\mathrm{h}$) and quasiparticle ($S^\mathrm{q}$, $C_V^\mathrm{q}$) phonon densities of states at 50, 100, 150, and 200~GPa, respectively.}
\label{fig-thermal}
\end{figure}

To examine temperature-dependent thermal properties of $Pm\bar{3}m$ NaH$_6$, we performed calculations at 50, 100, 150, and 200~GPa over temperatures from 20 to 300~K using a $2\times2\times2$ supercell. Figure~\ref{fig-thermal} shows the resulting mean-square displacements (MSDs), entropies, and heat capacities. At fixed pressure, the MSDs increase approximately linearly with temperature. At 220~K, the hydrogen MSD is 0.018~\AA$^2$ at 50~GPa and 0.008~\AA$^2$ at 200~GPa. The differences between thermodynamic quantities calculated from harmonic and quasiparticle phonon densities of states decrease with increasing pressure and are negligible at 200~GPa within this analysis.

\subsection*{S4. Competing NaH$_6$ phases at 50~GPa}

\begin{figure}[!htbp]
\centering
\begin{overpic}[width=0.9\columnwidth]{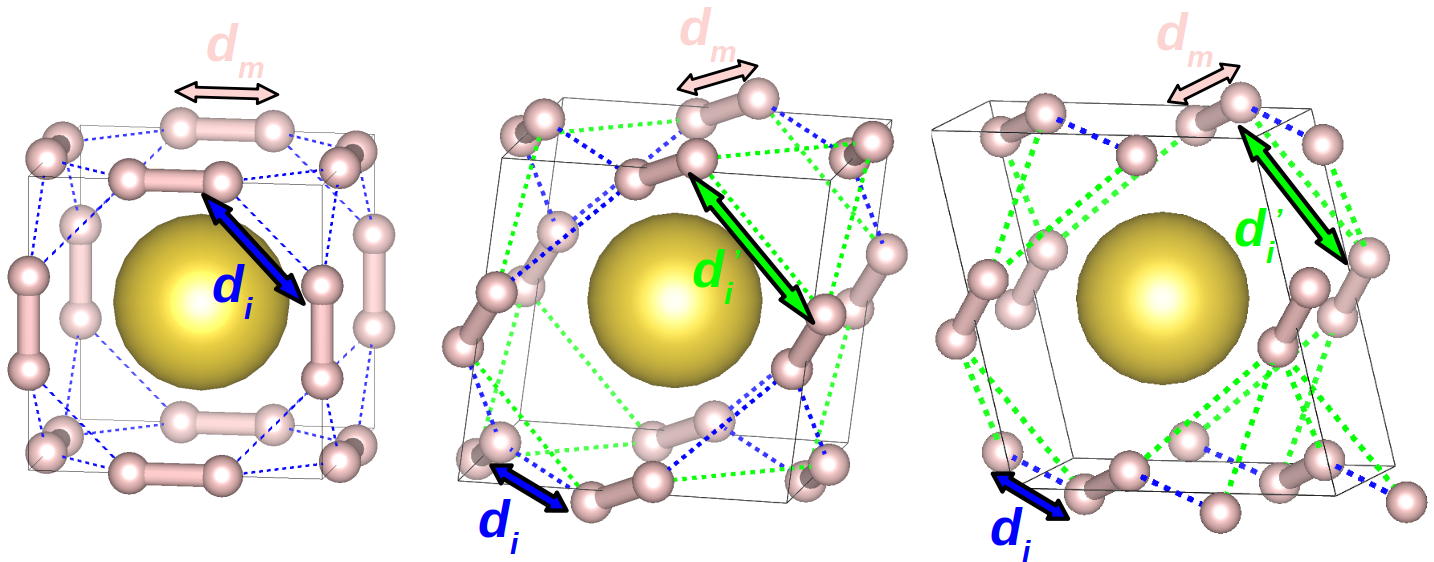}
\put (1,33) {\large\textbf a}
\put (34,33) {\large\textbf b}
\put (67,33) {\large\textbf c}
\end{overpic}
\caption{Crystal structures of (a) $Pm\bar{3}m$, (b) $R\bar{3}m$, and (c) $C2/m$ NaH$_6$ at 50~GPa.}
\label{fig-3phases}
\end{figure}

The crystal structures of $Pm\bar{3}m$, $R\bar{3}m$, and $C2/m$ NaH$_6$ at 50~GPa are compared in Fig.~\ref{fig-3phases}. The physical properties of the 15 metallic, dynamically stable NaH$_6$ structures identified at this pressure are listed in Table~\ref{table-15phase}. Among these structures, P3, P10, and P13 correspond to the $R\bar{3}m$, $C2/m$, and $Pm\bar{3}m$ phases discussed in the main text.

\begin{table}[!htbp]
\caption{Metallic, dynamically stable NaH$_6$ structures identified at 50~GPa. The ``higher symmetry'' and ``tolerance'' columns report the higher-symmetry structure assigned by structural matching and the corresponding matching tolerance. Bold entries P3, P10, and P13 denote the representative $R\bar{3}m$, $C2/m$, and $Pm\bar{3}m$ structures discussed in the main text. These structures serve as the reference structures for the structural-matching analysis; consequently, no higher-symmetry match or tolerance is listed for these rows. Volumes are per formula unit.}
\label{table-15phase}
\centering
\begin{tabular}{|c|c|c|c|c|c|}
\hline
ID & Enthalpy & Volume & Symmetry & Higher symmetry & Tolerance \\
 & (meV) & (\AA$^3$) & & & (\AA) \\
\hline
P1 & $-167.73$ & 27.491 & $P\bar{1}$ & $R\bar{3}m$ & 0.002 \\
P2 & $-167.73$ & 27.489 & $P\bar{1}$ & $R\bar{3}m$ & 0.002 \\
\textbf{P3} & $\bm{-167.71}$ & $\bm{27.491}$ & $\bm{R\bar{3}m}$ & & \\
P4 & $-167.70$ & 27.490 & $C2/m$ & $R\bar{3}m$ & 0.001 \\
P5 & $-167.70$ & 27.491 & $P\bar{1}$ & $R\bar{3}m$ & 0.005 \\
P6 & $-167.70$ & 27.491 & $P\bar{1}$ & $R\bar{3}m$ & 0.002 \\
P7 & $-167.69$ & 27.491 & $P\bar{1}$ & $R\bar{3}m$ & 0.003 \\
P8 & $-167.69$ & 27.491 & $P\bar{1}$ & $R\bar{3}m$ & 0.002 \\
P9 & $-162.45$ & 27.469 & $C2/m$ & $R\bar{3}m$ & 0.011 \\
\hline
\textbf{P10} & $\bm{-130.18}$ & $\bm{27.47}$ & $\bm{C2/m}$ & & \\
\hline
P11 & $-4.36$ & 26.804 & $P4/mmm$ & $Pm\bar{3}m$ & 0.0002 \\
P12 & $-4.14$ & 26.804 & $P\bar{1}$ & $Pm\bar{3}m$ & 0.013 \\
\textbf{P13} & $\bm{0.00}$ & $\bm{26.787}$ & $\bm{Pm\bar{3}m}$ & & \\
P14 & 0.26 & 26.787 & $Cmmm$ & $Pm\bar{3}m$ & 0.015 \\
P15 & 1.86 & 26.782 & $Cmmm$ & $Pm\bar{3}m$ & 0.010 \\
\hline
\end{tabular}
\end{table}

\subsection*{S5. $R\bar{3}m$ NaH$_6$: superconducting properties and finite-temperature stability}

\begin{figure}[!htbp]
\centering
\begin{overpic}[width=0.5\columnwidth]{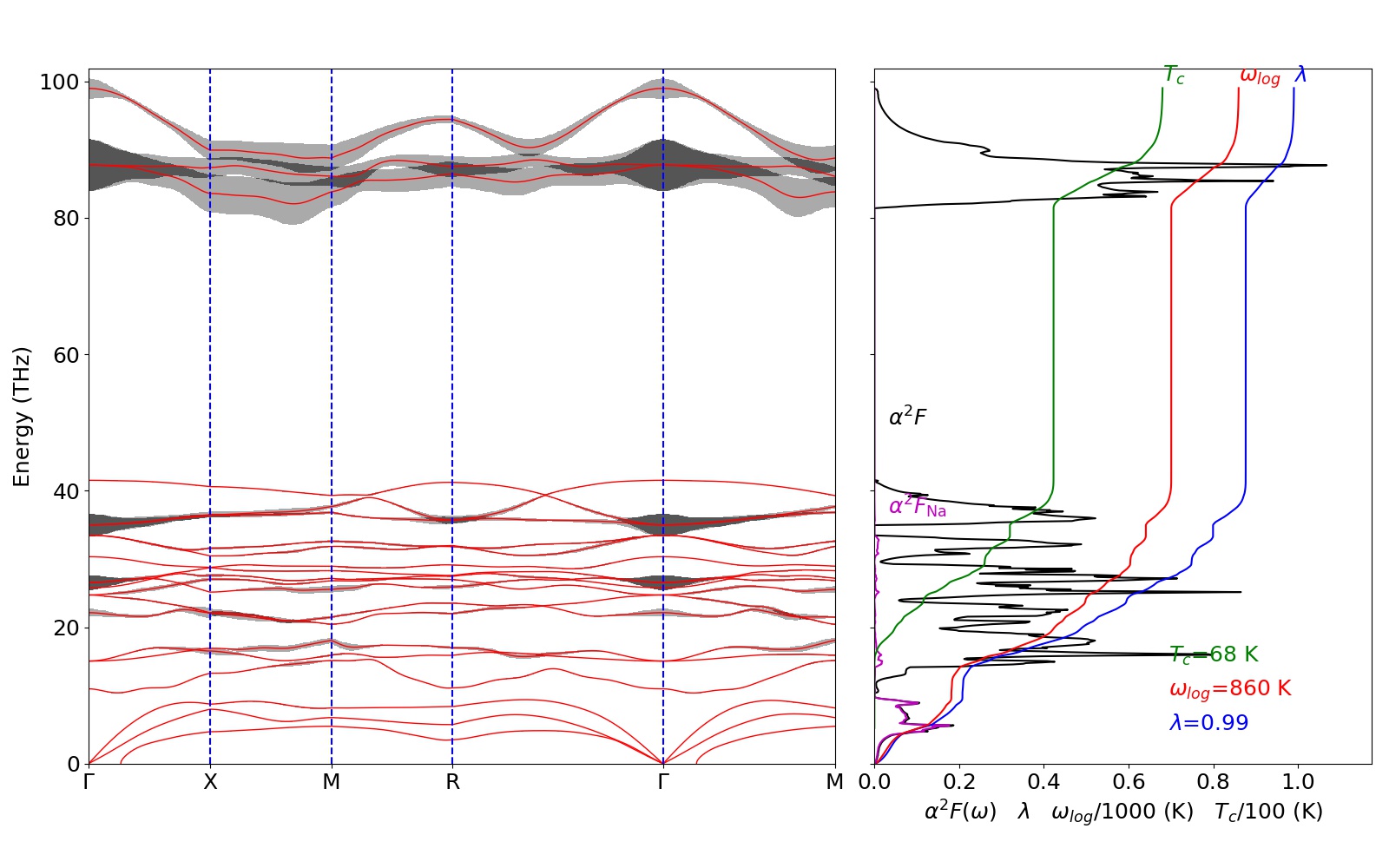}
\put (0,58) {\large\textbf a}
\put (63,58) {\large\textbf b}
\end{overpic}
\begin{overpic}[width=0.5\columnwidth]{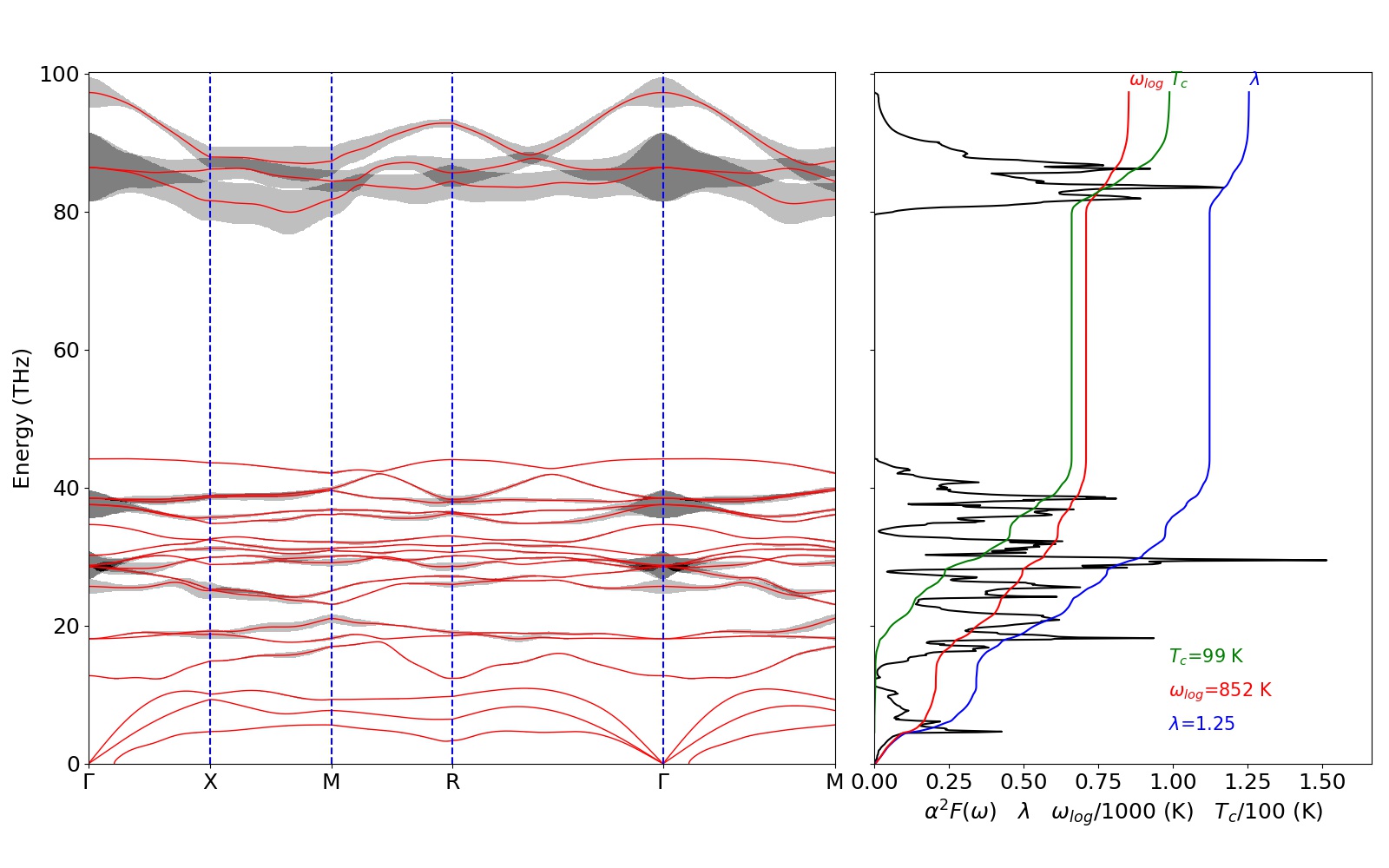}
\put (0,58) {\large\textbf c}
\put (63,58) {\large\textbf d}
\end{overpic}
\begin{overpic}[width=0.5\columnwidth]{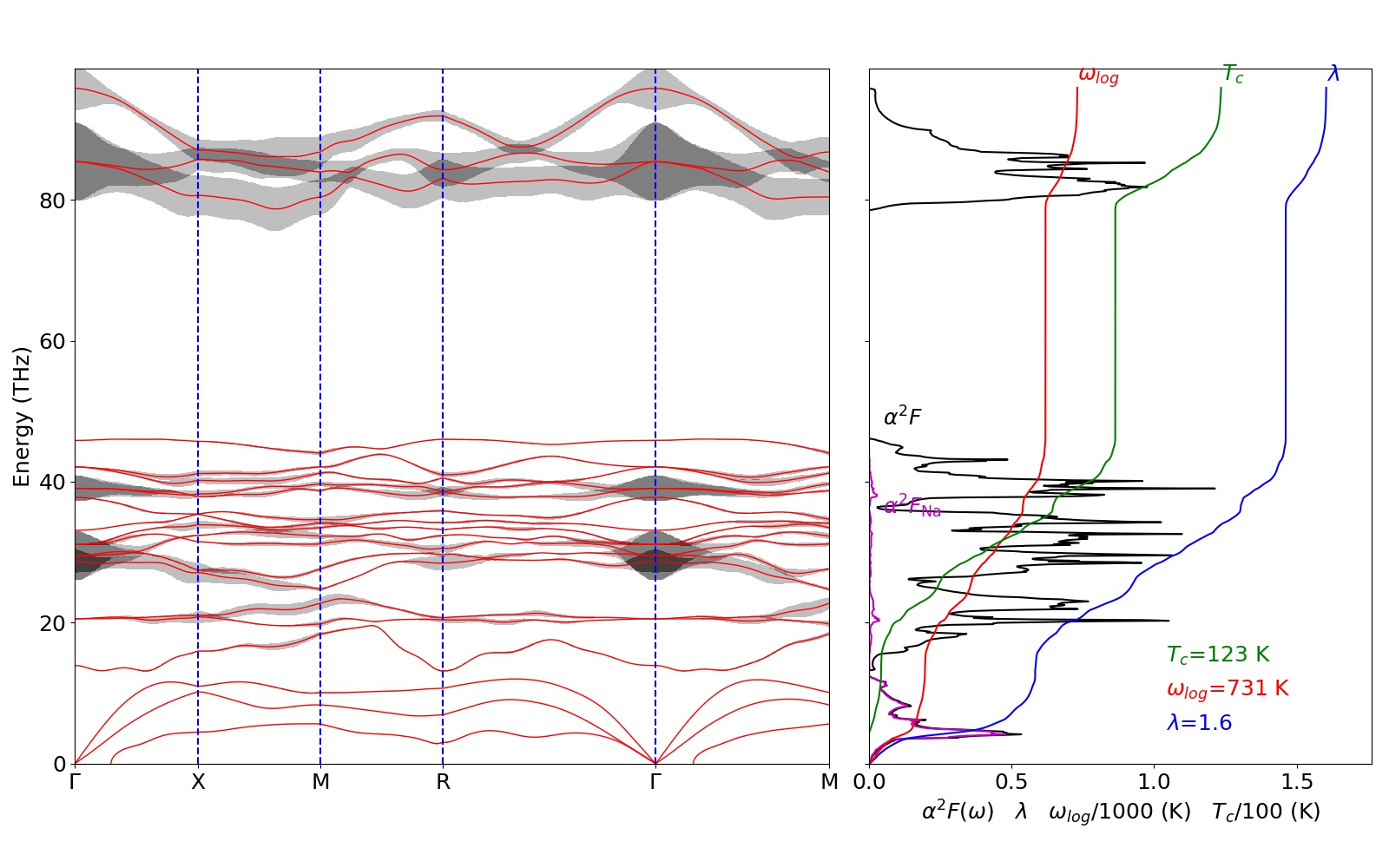}
\put (0,58) {\large\textbf e}
\put (63,58) {\large\textbf f}
\end{overpic}
\begin{overpic}[width=0.5\columnwidth]{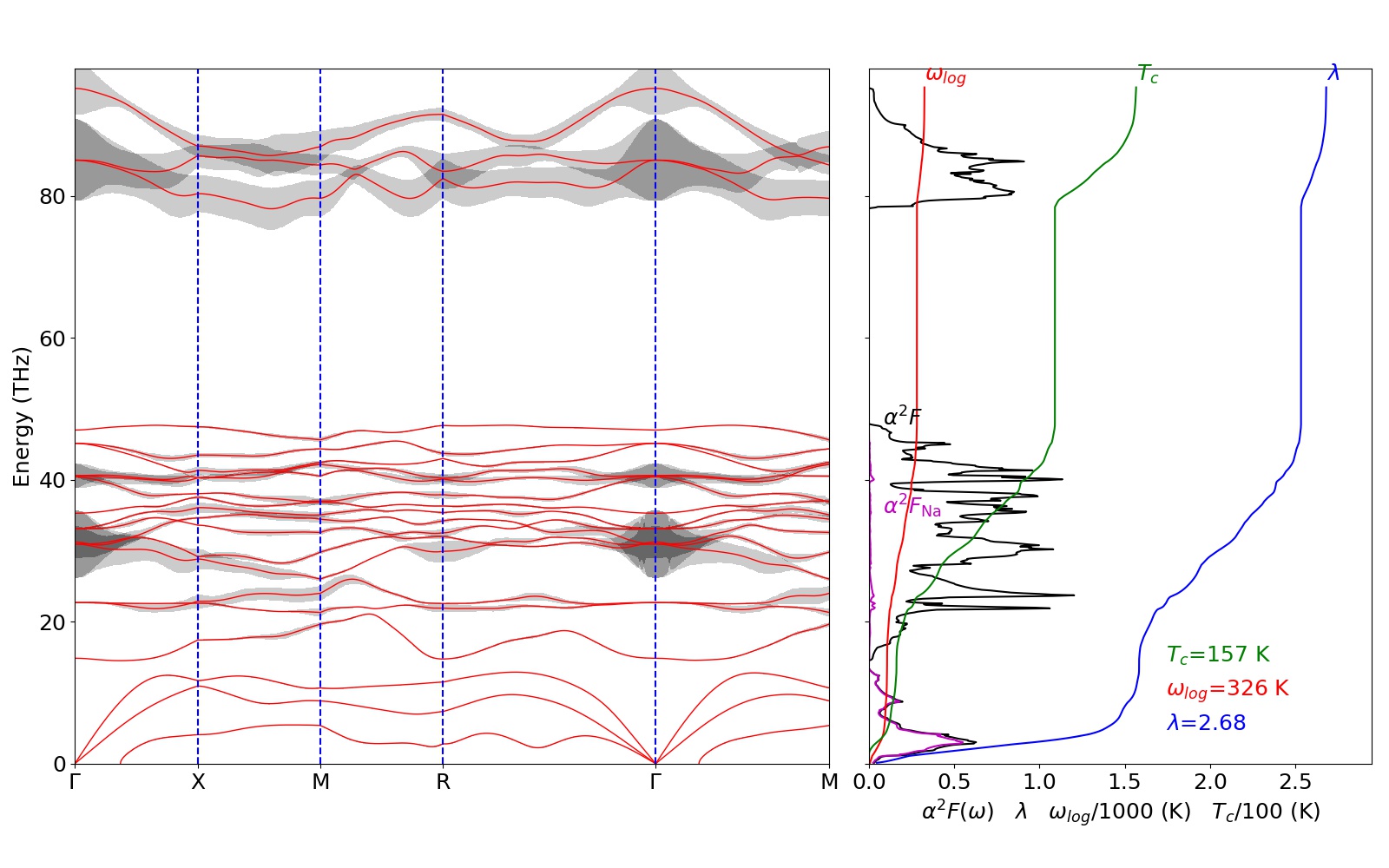}
\put (0,58) {\large\textbf g}
\put (63,58) {\large\textbf h}
\end{overpic}
\caption{(a), (c), (e), and (g) Phonon dispersions of $R\bar{3}m$ NaH$_6$ along $\Gamma$--$X$--$M$--$R$--$\Gamma$--$M$ at 30, 50, 70, and 90~GPa, respectively. The gray shaded areas represent the phonon linewidths. (b), (d), (f), and (h) Total and Na-projected $\alpha^2F(\omega)$, together with the integrated $T_\mathrm{c}$, $\omega_\mathrm{log}$, and $\lambda$, at the corresponding pressures. The black and purple curves show the total and Na-projected $\alpha^2F(\omega)$, respectively; the green, red, and blue curves show the integrated $T_\mathrm{c}$, $\omega_\mathrm{log}$, and $\lambda$, respectively.}
\label{fig-a2FR-3mApp}
\end{figure}

The phonon dispersions, $\alpha^2F(\omega)$, $\lambda$, $\omega_\mathrm{log}$, and $T_\mathrm{c}$ of $R\bar{3}m$ NaH$_6$ at 30, 50, 70, and 90~GPa are shown in Fig.~\ref{fig-a2FR-3mApp}, and the corresponding $T_\mathrm{c}$, $\omega_\mathrm{log}$, and $\lambda$ are listed in Table~\ref{table-TcR-3m}. The $T_\mathrm{c}$ of $R\bar{3}m$ NaH$_6$ increases from 34~K at 10~GPa to 157~K at 90~GPa.

\begin{table}[!htbp]
\caption{Calculated superconducting parameters of $R\bar{3}m$ NaH$_6$ ($\mu^*=0.10$).}
\label{table-TcR-3m}
\centering
\begin{tabular}{|c|c|c|c|}
\hline
$P$ (GPa) & $T_\mathrm{c}$ (K) & $\omega_\mathrm{log}$ (K) & $\lambda$ \\
\hline
10 & 34 & 797 & 0.73 \\ 
30 & 68  & 860 & 0.99 \\
50 & 99  & 852 & 1.25 \\
70 & 123 & 731 & 1.60 \\
90 & 157 & 326 & 2.68 \\ 
\hline
\end{tabular}
\end{table}

\begin{figure}[!htbp]
\centering
\includegraphics[width=0.8\columnwidth]{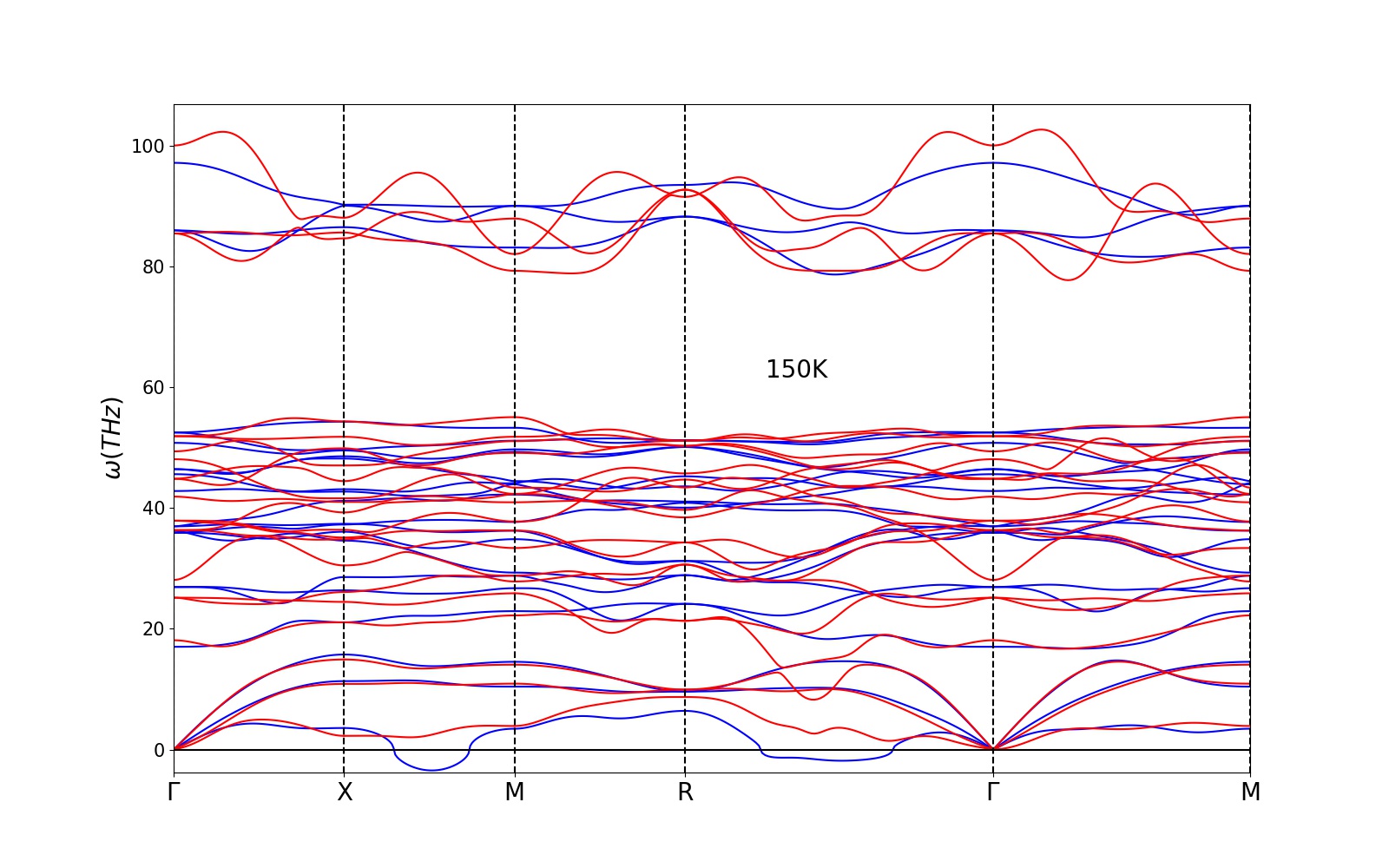}
\caption{Harmonic (blue) and renormalized (red) phonon dispersions of $R\bar{3}m$ NaH$_6$ at 110~GPa and 150~K.}
\label{fig-renormR-3m}
\end{figure}

The renormalized phonon dispersions of $R\bar{3}m$ NaH$_6$ at 110~GPa and 150~K, extracted from molecular-dynamics simulations using a $5\times5\times5$ supercell, are shown as red curves in Fig.~\ref{fig-renormR-3m}. Harmonic phonon dispersions are shown in blue. Within this quasiparticle analysis, the renormalized branches are positive, whereas the harmonic calculation has imaginary frequencies along the $X$--$M$ and $R$--$\Gamma$ paths.

\subsection*{S6. Structural parameters}

The structural parameters of $Pm\bar{3}m$ and $R\bar{3}m$ NaH$_6$ are listed in Tables~\ref{table-parameterPm-3m} and \ref{table-parameterR-3m}, respectively.

\begin{table}[!htbp]
\caption{Structural parameters of $Pm\bar{3}m$ NaH$_6$.}
\label{table-parameterPm-3m}
\centering
\begin{tabular}{|c|c|c|c|c|c|c|}
\hline
$P$ (GPa) & $c$ (\AA) & Atom & $x$ & $y$ & $z$ & Wyckoff site \\
\hline
\multirow{2}{*}{50} & \multirow{2}{*}{2.992} & Na & 0.00000 & 0.00000 & 0.00000 & 1a \\
 & & H & 0.14368 & 0.50000 & 0.50000 & 6f \\ \hline
\multirow{2}{*}{100} & \multirow{2}{*}{2.7899} & Na & 0.00000 & 0.00000 & 0.00000 & 1a \\
 & & H & 0.15130 & 0.50000 & 0.50000 & 6f \\ \hline
\multirow{2}{*}{150} & \multirow{2}{*}{2.6672} & Na & 0.00000 & 0.00000 & 0.00000 & 1a \\
 & & H & 0.15531 & 0.50000 & 0.50000 & 6f \\ \hline
\multirow{2}{*}{200} & \multirow{2}{*}{2.5794} & Na & 0.00000 & 0.00000 & 0.00000 & 1a \\
 & & H & 0.15786 & 0.50000 & 0.50000 & 6f \\ \hline
\end{tabular}
\end{table}

\begin{table}[!htbp]
\caption{Structural parameters of $R\bar{3}m$ NaH$_6$. Atomic coordinates are given in the rhombohedral primitive cell; Wyckoff labels refer to the hexagonal setting.}
\label{table-parameterR-3m}
\centering
\begin{tabular}{|c|c|c|c|c|c|c|c|}
\hline
$P$ (GPa) & $c$ (\AA) & $\alpha$ ($^\circ$) & Atom & $x$ & $y$ & $z$ & Wyckoff site \\
\hline
\multirow{2}{*}{10} & \multirow{2}{*}{3.6098} & \multirow{2}{*}{74.423} & Na & 0.00000 & 0.00000 & 0.00000 & 3a \\
 & & & H & 0.49625 & 0.50375 & 0.05250 & 18h \\ \hline
\multirow{2}{*}{30} & \multirow{2}{*}{3.2485} & \multirow{2}{*}{77.228} & Na & 0.00000 & 0.00000 & 0.00000 & 3a \\
 & & & H & 0.48885 & 0.51115 & 0.05981 & 18h \\ \hline
\multirow{2}{*}{50} & \multirow{2}{*}{3.072} & \multirow{2}{*}{78.67} & Na & 0.00000 & 0.00000 & 0.00000 & 3a \\
 & & & H & 0.48358 & 0.51642 & 0.06330 & 18h \\ \hline
\multirow{2}{*}{70} & \multirow{2}{*}{2.9573} & \multirow{2}{*}{79.743} & Na & 0.00000 & 0.00000 & 0.00000 & 3a \\
 & & & H & 0.47930 & 0.52070 & 0.06546 & 18h \\ \hline
\multirow{2}{*}{90} & \multirow{2}{*}{2.8697} & \multirow{2}{*}{80.695} & Na & 0.00000 & 0.00000 & 0.00000 & 3a \\
 & & & H & 0.47550 & 0.52450 & 0.06688 & 18h \\ \hline
\end{tabular}
\end{table}